\documentclass[11pt,a4paper]{article}
\usepackage{jheppub}

\usepackage{graphicx}
\usepackage{epsfig}
\usepackage{bm}
\usepackage{color}
\usepackage[utf8]{inputenc} 
\usepackage[T1]{fontenc}

\usepackage{booktabs} 
\usepackage{array} 
\usepackage{verbatim} 
\usepackage{slashed}
\usepackage[compat=1.1.0]{tikz-feynman}
\usepackage{hyperref}
\usepackage{caption}
\usepackage{subcaption}

\numberwithin{equation}{section}

\definecolor{DGreen}{rgb}{0.08,0.5,0.15}

\newcommand{\bit}{\begin{itemize}}
\newcommand{\eit}{\end{itemize}}
\newcommand{\ben}{\begin{enumerate}}
\newcommand{\een}{\end{enumerate}}
\newcommand{\beq}{\begin{equation}}
\newcommand{\eeq}{\end{equation}}
\newcommand{\bea}{\begin{eqnarray}}
\newcommand{\eea}{\end{eqnarray}}
\newcommand{ \lsim}{\mathrel{\vcenter
     {\hbox{$<$}\nointerlineskip\hbox{$\sim$}}}}
\newcommand{ \gsim}{\mathrel{\vcenter
     {\hbox{$>$}\nointerlineskip\hbox{$\sim$}}}}
\newcommand{\gappeq}{\mathrel{\rlap {\raise.5ex\hbox{$>$}}
{\lower.5ex\hbox{$\sim$}}}}
\newcommand{\lappeq}{\mathrel{\rlap{\raise.5ex\hbox{$<$}}
{\lower.5ex\hbox{$\sim$}}}}

\newcommand{\LNP}{\Lambda_{NP}}

\newcommand{\me}{\mu \! \to \! e}
\newcommand{\muc}{\mu A \to \! eA }
\newcommand{\mec}{\mu \! \to \! e~ {\rm conversion}}

\newcommand{\meg}{\mu \to e \gamma}

\newcommand{\meee}{\mu \to e \bar{e} e}

\def\a{\alpha}
\def\b{\beta}
\def\g{\gamma}
\def\d{\delta}

\def\m{\mu}

\def\r{\rho}
\def\s{\sigma}

\title{Distinguishing models with $\mu \to e$ observables}
\author[a,b]{Marco Ardu}
\author[a]{Sacha Davidson}
\author[c]{and Stéphane Lavignac }
\affiliation[a]{LUPM, CNRS,
	Université Montpellier\\
	Place Eugene Bataillon, F-34095 Montpellier, Cedex 5, France}
\affiliation[b]{Departament de Física Tèorica, Universitat de València,\\
	Dr. Moliner 50, E-46100 Burjassot (València), Spain}
\affiliation[c]{Institut de Physique Th{\'e}orique, Universit{\'e} Paris Saclay,\\
	CNRS, CEA, F-91191 Gif-sur-Yvette, France}
\emailAdd{marco.ardu@umontpellier.fr}
\emailAdd{s.davidson@lupm.in2p3.fr}
\emailAdd{stephane.lavignac@ipht.fr}
\abstract{Upcoming experiments will improve the reach for  the lepton flavour violating (LFV) processes $\mu \to e \gamma$, $\mu \to e \bar{e} e$ and $\mu A \to e A$ by orders of magnitude.
	We investigate whether this upcoming  data could rule out some popular TeV-scale LFV models (the type II seesaw, the inverse seesaw  and a scalar  leptoquark)
	using a bottom-up EFT approach. We take the data to be the twelve Wilson coefficients that  experiments can constrain and 
	in principle  determine independently.
	In this 12-dimensional  coefficient space,  each model  can only predict points in a  specific subspace; for instance, flavour change involving singlet electrons is suppressed in the seesaw models, and the leptoquark induces negligible  coefficients for  4-lepton scalar operators. 
	Using the fact that none of these models can populate the whole region accessible
	to upcoming experiments, we show that $\mu \to e$ experiments have the ability
	to rule them out.}

\begin{document}
\maketitle


\section{Introduction and review}
\label{intro}

\subsection{Introduction}

Searches for New Physics(NP) in the lepton sector  are of great interest,
because  such NP is required by  neutrino masses,   it could fit some   current anomalies (such as $(g-2)_\mu$~\cite{Muong-2:2021ojo} and observations in $B$ meson physics~\cite{BaBar:2012obs,Belle:2015qfa,Belle:2019rba,LHCb:2023zxo,LHCb:2023cjr}), and  because leptons do not have  strong interactions, so the observables are relatively  clean.  In this paper, we assume  that this  leptonic  New Physics  is heavy,  and
  parametrise it in EFT~\cite{Georgi,burashouches,LesHouches}. 

Lepton Flavour change (LFV) in the  $\me$ sector is   promising for  the discovery of leptonic NP, because the    experimental sensitivity is  already good, and  is expected to improve by several orders of magnitude in the near future
  (see table \ref{tab:bds}).  
  However,  few processes are constrained, so the current  experimental bounds only set about a dozen constraints on Wilson coefficients~\cite{DKY}.  One  can therefore  wonder  whether future   observations of  $\me$ flavour change  could  distinguish among  the multitude of models that predict LFV.

\begin{table}[ht]
\begin{center}
 \begin{tabular}{|l|l|l|}
 \hline
 process & current bound & future reach \\
\hline
$\meg $ & $ < 4.2 \times 10^{-13}$(MEG~\cite{TheMEG:2016wtm})
 &$6 \times 10^{-14}$ (MEG II~\cite{MEGII}) \\
$\meee $& $ < 1.0 \times 10^{-12}$(SINDRUM~\cite{Bellgardt:1987du}) 
 & $ \sim 10^{-16}$ (Mu3e~\cite{Mu3e}) \\
$\mu$Au$ \to e$Au & 
$< 7 \times 10^{-13}$(SINDRUM II~\cite{Bertl:2006up}) &
? $ \to 10^{-(18\to 20)}$ (PP/AMF~\cite{PP,CGroup:2022tli})  \\
$\mu$Ti$ \to e$Ti & 
$< 6.1 \times 10^{-13}$(SINDRUM II~\cite{Wintz:1998rp}) &
 $ \sim 10^{-16}$ (COMET~\cite{COMET}, Mu2e~\cite{mu2e})  \\
\hline
$\tau\to l + \dots$ & $\lsim  10^{-8}$(Babar/Belle)~\cite{tau1,Belle:2007cio}
& $ \sim 10^{-(9\to 10)} $(BelleII)~\cite{belle2t3l} \\
\hline
 \end{tabular}
 \caption{Current bounds on the branching ratios for various lepton flavour changing processes, and estimated reach 
 of   ``upcoming'' experiments, i.e. those under construction or running, { as well as of the proposals PRISM/PRIME (PP)
 and Advanced Muon Facility (AMF).}} \label{tab:bds} 
 \end{center}
 \end{table}

Predictions  for $\me$  LFV have been widely studied over several decades in a multitude of models, such as the supersymmetric type I seesaw, the supersymmetric type II seesaw, supersymmetric flavour models, {left-right symmetric models}, two Higgs doublet models, the inverse seesaw and its supersymmetric version, warped extra dimensions, the littlest Higgs model with T parity, unparticle physics, radiative neutrino mass models, spontaneous lepton number violation, low-scale flavour models, and many others
(see e.g. Refs.~\cite{Hisano:1995cp,Antusch:2006vw,Rossi:2002zb,Altmannshofer:2009ne,Cirigliano:2004mv,Omura:2015nja,Arganda:2014dta,Deppisch:2004fa,Agashe:2006iy,Blanke:2007db,Aliev:2007qw,Cai:2017jrq,Escribano:2021uhf,Lopez-Ibanez:2021yzu}, and for recent reviews Refs.~\cite{Calibbi:2017uvl,Ardu:2022sbt}).
Top-down analyses -- which start from the model to predict  observables -- frequently show   correlations among branching ratios, often resulting from scans over model parameter space. In our bottom-up EFT perspective, starting from the data, we
address  a different question:  can  observations  distinguish among models?

In this paper, we  focus on  three  models  with new heavy particles around the TeV scale. The   first two are neutrino mass models : the TeV-scale version of the type II seesaw mechanism~\cite{Magg:1980ut,Lazarides:1980nt,Schechter:1980gr,Mohapatra:1980yp} and the inverse type I seesaw~\cite{Wyler:1982dd,Mohapatra:1986aw,Mohapatra:1986bd}, {whose predictions for LFV processes have been studied, mainly in the top-down approach, by many authors (see e.g. Refs.~\cite{Chun:2003ej,Kakizaki:2003jk,Akeroyd:2009nu,Dinh:2012bp,Barrie:2022ake} for the type II}
{and Refs.~\cite{Tommasini:1995ii,Ilakovac:1994kj,Ibarra:2011xn,Dinh:2012bp,Alonso:2012ji,Abada:2015oba,Coy:2018bxr,Abada:2021zcm,Zhang:2021jdf,Coy:2021hyr,Granelli:2022eru,Crivellin:2022cve} for the inverse seesaw, where~\cite{Coy:2018bxr,Zhang:2021jdf,Coy:2021hyr,Crivellin:2022cve} follow an EFT approach).}
Both these models have the additional attraction of  being able to generate the baryon asymmetry of the Universe via leptogenesis~\cite{Fukugita:1986hr} (for a review, see Ref.~\cite{Davidson:2008bu}).
While, in the type II seesaw case, thermal leptogenesis requires a triplet mass above $10^{10}\, {\rm GeV}$ or so~\cite{Hambye:2005tk,AristizabalSierra:2014nzr,Lavignac:2015gpa}, a TeV-scale scalar triplet with non-minimal coupling to gravity can lead to successful leptogenesis~\cite{Barrie:2021mwi} through the Affleck-Dine mechanism~\cite{Affleck:1984fy}.
The inverse seesaw model, on the other hand, features TeV-scale sterile neutrinos which can generate the baryon asymmetry of the Universe through resonant leptogenesis~\cite{Pilaftsis:2003gt,Blanchet:2009kk,BhupalDev:2014pfm,daSilva:2022mrx}
or ARS leptogenesis~\cite{Akhmedov:1998qx,Asaka:2005pn,Klaric:2021cpi,Hernandez:2022ivz}.
The last model  is an SU(2) singlet leptoquark which can fit the $R_D$ anomaly~\cite{BaBar:2012obs,Belle:2015qfa,Belle:2019rba,LHCb:2023zxo,LHCb:2023cjr}, as discussed by many authors (see e.g. Refs.~\cite{Sakaki:2013bfa,Cai:2017wry,Angelescu:2018tyl,Lee:2021jdr}).
  The leptoquark differs from the   neutrino mass models  in that at tree level,  it generates
   2lepton-2quark operators (which mediate $\mec$), and in that  it couples  to  SU(2) singlet fermions of the SM.  

 We apply bottom-up EFT to explore  whether $\me$ LFV  can distinguish among models, starting from the observation that  the  data
 could determine 12 operator  coefficients, and not just the three branching ratios.  
 We consider this 12-dimensional coefficient space, and    ask whether the volume accessible to upcoming experiments can be filled by each of three models. So we aim to identify the region of the ellipse that a model cannot occupy; an observation in this region would rule the model out.
 Our study is performed in an EFT framework inspired by Ref.~\cite{C+C}, and  differs from  top-down analyses,  in that we do not scan over model parameter space for which  we do not known the measure, and because  we take the data to be 12 Wilson coefficients.  A more complete analysis and technical details will appear in a subsequent publication~\cite{pl2}.

Our EFT framework is briefly summarised  in the next subsection.
In the following three sections, we present  and discuss   three  models of New Physics at the TeV scale: the type II seesaw  in Section~\ref{sec:t2},  the   inverse  type I seesaw in Section~\ref{sec:t1}, and  a leptoquark in Section~\ref{sec:lq}. Section~\ref{sec:summ}  compares the models and summarises the results.

\subsection{Review}

We consider  three  processes --  $\meg$, $\meee$ and  Spin-Independent\footnote{Spin-Dependent $\muc$~\cite{SDmuc,Hoferichter:2022mna} is also possible, but analogously to WIMP scattering, is relatively suppressed.}  (SI)$\muc$ --  because they are complementary~\cite{C+C}, and because the experimental sensitivity could improve significantly in coming years. The branching ratios  are given in Appendix \ref{app:BRs},  in terms of the coefficients $\{C\}$
 of   the Lagrangian~\cite{KO} at the experimental scale:
\bea
\d {\cal L} &=& \frac{1}{v^2} \sum_{X\in L,R} {\Big [}
 C^{e \mu}_{D,X} (m_\mu \overline{e} \sigma^{\r\s}P_{X} \mu) F_{\r\s}
 + C^{e \mu e e}_{S,XX} (\overline{e} P_X \mu )(\overline{e} P_X e )
 + C^{e \mu e e}_{VXR} (\overline{e} \gamma^\r P_X \mu ) (\overline{e} \gamma_\r P_R e )
\nonumber\\
&& + C^{e \mu e e}_{VXL} (\overline{e} \gamma^\r P_X \mu ) (\overline{e} \gamma_\r P_L e )
  + C_{Alight,X} {\cal O}_{Alight,X} + C_{Aheavy,X} {\cal O}_{Aheavy,X} {\Big ]} +{\rm h.c}
\label{L1}
\eea
where the   twelve  $C$s are dimensionless complex numbers,
$X \in\{ L,R\}$, $\frac{1}{v^2} = 2\sqrt{2} G_F$ (so $v=$174 GeV), and
${\cal O}_{Alight,X}$  and $ {\cal O}_{Aheavy,X}$ are  respectively the four-fermion operator combinations that induce $\muc$ on light nuclei like Titanium or Aluminium, and  an  operator combination probed by heavy targets like Gold.
Expressions for these operators  are given in  Appendix \ref{app:mucops}.

\begin{table}[t]
 \begin{tabular}{|l|c|c|l|}
 \hline
 coefficient &  current bound &  future bound & process \\
\hline
$C_{D,X}^{e\mu}$ & $1.0\times 10^{-8}$ & $ \sim 10^{-9}$&$\meg,\meee $\\
$C_{V,XX}^{e\mu ee}$ & $0.7\times 10^{-6}$ & $\sim 10^{-8}$&$\meee $\\
$C_{V,XY}^{e\mu ee}$ & $1.0\times 10^{-6}$ & $\sim 10^{-8}$&$\meee $\\
$C_{S,XX}^{e\mu ee}$ & $2.8\times 10^{-6} $ & $\sim 10^{-8}$&$\meee $\\
$C_{Alight,X}$ & $5.0\times 10^{-8}$ & $\sim 10^{-10}$&$\mu {\rm Ti}\to e   {\rm Ti}$\\
$C_{Aheavy\perp,X}$ & $2\times 10^{-7}$ & &$\mu {\rm Au}\to e   {\rm Au}$\\
\hline
 \end{tabular}
 \caption{Current bounds on the operator coefficients of the Lagrangian~(\ref{L1}) at the experimental scale $m_\mu$ ($X = L, R$), and estimated reach of upcoming experiments { (not including the proposals PP and AMF).}}
  \label{tab:Cbds}
 \end{table}

The non-observation of  $\me$ processes constrains the  coefficients    in Eq.~(\ref{L1}) to sit in a 12-dimensional ellipse at the origin~\cite{DKY}. 
The counting of constraints and the  bounds obtained from the  correlation matrix  for $\meg$ and $\meee$ are discussed in ~\cite{DKY}; these results give   the current bounds, and the estimated sensitivities of upcoming experiments, listed in Table~\ref{tab:Cbds}.
  Observations could in principle  determine the magnitude  of    each   coefficient:  if the  decaying muon  is polarised~\cite{KO} (which can also be possible for $\muc$~\cite{Kadono:1986zz,Kuno:1987dp}) then  the  chirality of the $\mu\to e$ bilinear can be determined, and  asymmetries and angular distributions in $\meee$ can distinguish among most of  the four-lepton operators that contribute~\cite{Okadameee}.
  (Scalar ${\cal O}_{SXX}$ and vector ${\cal O}_{VYY}$ operators, for $X\neq Y$, induce the same angular distribution, but are distinguishable  via the  $e^\pm$ helicities \footnote{We thank  Ann-Kathrin Perrevoort of Mu3e for discussions. Also the scalar operators could be difficult to obtain in models.})
Some relative phases can also be measured~\cite{Okadameee}. Finally, changing the target  material in $\muc$ allows to probe different  combinations of  vector and scalar  coefficients on protons or neutrons~\cite{KKO,Haxton:2022piv}; current theoretical accuracy allows to obtain independent information from  at least two  targets~\cite{DKY,Davidson:2022nnl}, so  in this paper we   focus on light targets like  Titanium (used by SINDRUMII~\cite{Bertl:2006up,Wintz:1998rp}) or Aluminium (the target for the upcoming Mu2e and COMET experiments).  The complementary constraints that can be obtained with  Gold (used by SINDRUMII~\cite{Bertl:2006up}), will be discussed in~\cite{pl2}.
With theoretical optimism, we assume the coefficients can be distinguished  to the  reach of upcoming experiments.

We take  the New Physics scale  $\LNP \sim $ TeV  for  the three models considered here.
  The coefficients are evolved from the experimental scale $\sim m_\mu$ to $\LNP \sim$ TeV in the broken electroweak theory,  using the ``Leading Order''  RGEs  of QED and QCD~\cite{Crivellin:2017rmk,C+C} (starting respectively at   $m_\mu$ and  2 GeV), for the operator basis of Ref.~\cite{C+C}.
This  includes   the leading log-enhanced loops of QED and QCD via  the RGEs,
and loop diagrams with  the $W$,$Z$ or Higgs  can contribute in the matching.
We prefer this approach  over matching to SMEFT at the weak scale,  because 
it allows to resum QCD\footnote{It is convenient to  use 5 flavours at all scales, because the results  for 5  or 6 flavours are numerically similar.}  between the experimental scale and $\LNP$,   and  avoids the issue  that  $v$/TeV is not large, implying that the SMEFT expansions in $1/\LNP^2$ and $\a^n \ln$ may not
  converge quickly \footnote{This approximation may also double count   some
electroweak  contributions that we think are higher order, as  will be discussed  in~\cite{pl2}.}.
This RG evolution  gives  the  12-dimensional  ellipse at $\LNP$.  We then match onto each of the models in turn (at
  tree level in the EFT), and explore  whether they can fill the ellipse.

 {
In relating models to observables, 
it is convenient to  use as  stepping stones
the  coupling constant combinations that appear in Wilson coefficients, 
because  they parametrise the $\me$ LFV.  
For instance, tree level exchange of a leptoquark interacting with  $u$ quarks, matches onto  an  coefficient $\propto \lambda^{eu} \lambda^{\mu u *}$, and  the loop diagram of Fig.~\ref{fig:typeIIdipole} is $\propto  [f f^* Y_e]_{e\mu}$. We  refer to these combinations as ``invariants'' (a la Jarlskog), because they  are related  to $S$-matrix elements, and therefore  should  be independent of some Lagrangian redefinitions. 
}

The operator coefficients can of course be complex, and in some cases the relative phases are observable (for instance  in asymmetries  in $\meee$~\cite{Okadameee}). However, for plotting purposes, it is common to approximate the coefficients as real. In our analysis,  the coefficients  are  complex, but  we plot either the absolute values or the real parts;  the phases will be  discussed in~\cite{pl2}.


\section{ Type II seesaw }
\label{sec:t2}

  The first model we consider  is the type II seesaw mechanism~\cite{Magg:1980ut,Lazarides:1980nt,Schechter:1980gr,Mohapatra:1980yp},
  which  generates neutrino masses  via  the tree level exchange of an SU(2) triplet scalar $\Delta$.
 In this model, the Yukawa matrix is   directly proportional   to   the observed neutrino mass matrix,  so it is predictive  of flavour structure -- for instance fixing some  ratios between  $\tau \to l$ and $\mu \to e$  transitions -- and its  LFV signatures  have been widely studied~\cite{Chun:2003ej,Kakizaki:2003jk,Akeroyd:2009nu,Dinh:2012bp,Barrie:2022ake}.

 {We assume the  triplet  scalar is at the TeV scale, so could be produced at current and future colliders and lead to particular signatures~\cite{Akeroyd:2005gt,FileviezPerez:2008jbu,Melfo:2011nx, Freitas:2014fda,Ghosh:2017pxl,Antusch:2018svb}. It could also affect Higgs physics~\cite{Arhrib:2011uy,Arhrib:2011vc} and contribute to electroweak observables such as $m_W$~\cite{CDF:2022hxs}.}

  The SM Lagrangian is augmented by the following interactions
\bea
\d{\cal L}_\Delta&=& (D_\rho \Delta^I)^\dagger D^\rho \Delta^I  -M_\Delta^2|\Delta|^2
+\frac12\left(f_{\alpha\beta}\,\overline{\ell^c_\alpha} (i\tau_2)\tau_I \ell_\beta \Delta^I
+M_\Delta \lambda_H\, H^T (i\tau_2) \tau_I H \Delta^{*I}+{\rm h.c.}\right)
\nonumber          \\
&&     + \lambda_3 (H^\dagger H) (\Delta^{I*} \Delta^{I})+\lambda_4{\rm Tr}(\Delta^{I*}\tau_{I}\tau_{J}\tau_{K}\Delta^{K})(H^\dagger \tau_J H) + \dots ~~,
\label{LII}
\eea
where $\Delta$  is the colour-singlet, $SU(2)$-triplet  scalar of  hypercharge $Y_{\Delta}=+1$, $\ell$ are the left-handed SU(2) doublets,
 $M_\Delta$ is  the  triplet  mass which we take $\sim$ TeV,
$f$ is a symmetric complex $3\times3$ matrix  proportional to the light neutrino mass matrix and whose indices $
\a,\b$ run   over $\{e,\mu,\tau\}$,
$\{\tau_I\}$ are the Pauli matrices,
and the $\lambda$'s are real dimensionless couplings\footnote{$\lambda_H$ can be taken real with no loss of generality.}.
The dots on the right-hand side of Eq.~(\ref{LII}) stand for scalar interactions that are not relevant for LFV processes. {We also find negligible contributions to LFV operators from the triplet-Higgs interactions assuming perturbative $\lambda_{3,4}$. Consequently, these contributions will not be included in the subsequent discussion.}

  We match  the model to EFT at  the  scale $M_\Delta\sim $TeV,
  generating  a neutrino mass matrix $[m_\nu]_{\a\b} = U_{\a i}m_iU_{\b i}$   via   the tree-level exchange of the triplet between  pairs of leptons and Higgses:
\begin{equation}
[m_\nu]_{\alpha\beta}\simeq 0.03\ {\rm eV}\ f^*_{\alpha\beta} \frac{\lambda_H}{10^{-12}}\frac{{\rm TeV}}{M_{\Delta}} ~~.
\label{mnut2} 
\end{equation}

Exchanging the triplet  among four leptons matches onto   one of  the LFV
  coefficients of Eq.~(\ref{L1}), which induces $\meee$:
\bea
C^{e\mu ee}_{V,LL} \simeq \frac{v^2}{2 M^2_{\Delta}}f_{\mu e}f^*_{e e}
=\frac{[m^*_{\nu}]_{\mu e}[m_{\nu}]_{e e}}{2 \lambda_H^2v^2} ~~.
\label{CVLL}
\eea

 The   small ratio $m_\nu/v$  can  be obtained by suppressing  $\lambda_H$,   while leaving  unconstrained  $f v/M_\Delta$, which controls the magnitude of LFV. 
The triplet Yukawa matrix  $[f]$  is proportional to $[m_\nu]$, 
 so its flavour structure  can   be determined  from neutrino oscillation data~\cite{ParticleDataGroup:2022pth}. The only unknowns are  the mass $m_{min}$ of the  lightest neutrino,  two Majorana phases, and the Hierarchy (Normal = $m_3>m_2>m_1$, Inverted = $m_3<m_1< m_2$).  We use Eq.~(\ref{mnut2}) in order to  express  $[f]$  in terms of   $[m_\nu]$.

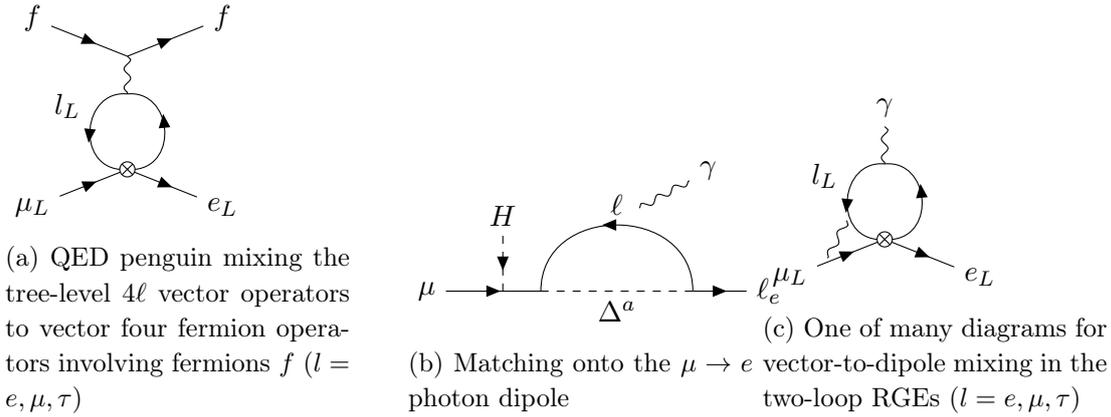
\begin{figure}
	\begin{subfigure}{0.3\textwidth}
		\begin{tikzpicture}
			\begin{feynman}[small]
				\vertex (mu) at (-1.25,-0.5) {\(\mu_L\)};
				\vertex (e) at (1.25,-0.5) {\(e_L\)};
				\vertex (b) at (0,1);
				\vertex (gamma) at (0,1.5);
				\vertex (f1) at (-1.25,2) {\(f\)};
				\vertex (f2) at (1.25,2) {\(f\)};
				\vertex[crossed dot] (a)  at (0,0) {};
				\diagram* [inline=(a.base)]{
					(mu) -- [fermion] (a),
					(a) -- [fermion] (e),
					(a) -- [anti fermion, half left, edge label=\(l_L\)] (b) -- [anti fermion, half left] (a),
					(b) -- [photon] (gamma),
					(f1) -- [fermion] (gamma) -- [fermion] (f2),
				};
			\end{feynman}
		\end{tikzpicture}
		\caption{QED penguin mixing the tree-level  4$\ell$ vector operators  to vector four fermion operators  involving fermions $f$ ($l=e,\mu,\tau$)}\label{fig:typeIIpenguin}
	\end{subfigure}\qquad 
  \begin{subfigure}{0.3\textwidth}
	\begin{tikzpicture}
		\begin{feynman}[small]
			\vertex (mu) at (-1.5,0) {\(\mu\)};
			\vertex (e) at (3,0) {\(\ell_e\)};
			\vertex (H1) at (-0.5,1) {\(H\)};
			\vertex (H2) at (-0.5,0);
			\vertex (a)  at (0,0) ;
			\vertex (b) at (2,0) ;
			\vertex (boson1) at (1.5, 1);
			\vertex (boson2) at (2, 2);
			\vertex (c) at (1.3,1.1) ;
			\vertex (H3) at (2.2,1.6) {\(\gamma\)};
			\diagram* [inline=(a.base)]{
				(H1) -- [charged scalar] (H2),
				(H3) -- [photon] (c),
				(mu) -- [fermion] (a),
				(b) -- [fermion] (e),
				(a) -- [anti fermion, half left, edge label=\(\ell\)] (b)  -- [scalar, edge label=\(\Delta^a\)] (a),
			};
		\end{feynman}
	\end{tikzpicture}
	\caption{Matching onto the $\mu\to e$ photon dipole }\label{fig:typeIIdipole}
  \end{subfigure}
\begin{subfigure}{0.3\textwidth}
	\begin{tikzpicture}
		\begin{feynman}[small]
			\vertex (mu) at (-1.25,-0.5) {\(\mu_L\)};
			\vertex (e) at (1.25,-0.5) {\(e_L\)};
			\vertex (b) at (0,1);
			\vertex (gamma) at (0,1.75) {\(\gamma\)};
			\vertex (gamma1) at (-0.75,-0.25);
			\vertex (gamma2) at (-0.5, 0.25);
			\vertex[crossed dot] (a)  at (0,0) {};
			\diagram* [inline=(a.base)]{
				(mu) -- [fermion] (a),
				(a) -- [fermion] (e),
				(a) -- [anti fermion, half left, edge label=\(l_L\)] (b) -- [anti fermion, half left] (a),
				(b) -- [photon] (gamma),
                (gamma1) -- [photon] (gamma2)
			};
		\end{feynman}
	\end{tikzpicture}
	\caption{One of many diagrams for vector-to-dipole mixing in the two-loop RGEs ($l=e,\mu,\tau$)}\label{fig:typeIItwoloopdipole}
\end{subfigure}
   \caption{ Loop contributions to $\mu \to e$
   operators and to their mixing in the type II seesaw model.}
\end{figure}

The type II seesaw will  also  induce
other  LFV  coefficients  given in the  Lagrangian (\ref{L1}).
Tree-level triplet exchange matches onto 4$\ell$ operators with $\mu$ and $\tau$
bilinears,  and these combine with Eq.~(\ref{CVLL}) in a ``penguin'', as
illustrated in Fig.~\ref{fig:typeIIpenguin},  to generate, for instance
\bea
C^{e\mu ee}_{V,LR}& =& \frac{\alpha_e}{3\pi}
\left[\frac{[m^\dagger_{\nu} m_{\nu}]_{\mu e }}{\lambda_H^2v^2}
  \ln \left(\frac{M_\Delta}{m_\tau}\right)
  +\sum_{\a \in e,\mu}\frac{[m^*_{\nu}]_{\mu \a}[m_{\nu}]_{e \a}}{\lambda_H^2 v^2}
  \ln\left(\frac{m_\tau}{m_\mu}\right)\right]
\label{ping}
\eea
This loop  arises in the RGEs, or equivalently, is log-enhanced. The logarithm is cut off at low energy by  the experimental scale ($m_\mu$) or the  mass of the lepton in the loop, so the $\tau$   is not included  in the loop between $m_\tau \to m_\mu$.  This is interesting, because $[m^\dagger_{\nu} m_{\nu}]_{\mu e }$,
which  appears in the first term of Eq.~(\ref{ping}) is determined by
neutrino oscillation parameters\footnote{Here and in the rest of this paper, we assume $\delta = 3 \pi/2$, a value consistent with the hints for CP violation in the lepton sector from the T2K experiment~\cite{T2K:2023smv}.} $$[m^\dagger_{\nu} m_{\nu}]_{\mu e } \sim i \sin\theta_{13} \Delta_{atm}^2~~,$$
so any  dependence of $C^{e\mu ee}_{V,LR}$  on the   unknown  neutrino mass scale  or Majorana phases can only arise from the second term. 
This same penguin diagram  also generates a loop correction to $C_{V, LL}^{e\mu ee}$  of  Eq.~(\ref{CVLL}), and contributes to $\muc$  on the proton
\bea
\Delta C^{e\mu ee}_{V,LL} =  C^{e\mu ee}_{V,LR}
~~~,~~~C^{e\mu pp}_{V,L} = - 2C^{e\mu ee}_{V,LR}
\label{ping2}
\eea
where $C^{e\mu pp}_{Alight,L} = \frac{1}{2}C^{e\mu pp}_{V,L} +...  $.  The
coefficient on neutrons, $C^{e\mu nn}_{V,L}$, vanishes at the order we calculate.

Finally, the dipole coefficients   are induced by   one loop  matching(see Fig.~\ref{fig:typeIIdipole}),  and  shrink marginally  in  running down  to the experimental scale, while being regenerated  
at two loop \footnote{The two-loop  diagrams~\cite{Ciuchini:1993fk,Degrassi:1998es,Czarnecki:2002nt,Crivellin:2017rmk} are included here because they are  ``leading order'' in the RGEs,
and  because  they are  numerically significant -- for instance, in the electroweak contribution to $(g-2)_\mu$, the log-enhanced 2-loop  contribution is $\sim$ 1/4 of the 1-loop matching part.}   as illustrated in Fig.~\ref{fig:typeIItwoloopdipole}:
\bea
  C^{e\mu}_{D,R}& \simeq &
  \frac{3e}{128 \pi^2}  {\Big [}\frac{[m_\nu m_\nu^\dagger]_{e\mu}}{\lambda^2_H v^2}
  {\Big (}1  + \frac{32}{27}  \frac{\alpha_e}{4\pi} \ln\frac{M_\Delta}{m_\tau} 
    {\Big )}   +   
  \frac{116 \alpha_e}{27\pi }  \ln\frac{m_\tau}{m_\mu} \sum_{\a \in e\mu} \frac{[m_\nu]_{\mu \a} [m^*_\nu]_{e\a} }{\lambda_H^2 v^2 }{\Big ]}
\label{CD}
\eea
where the  first (leading) term is independent of the neutrino mass scale and Majorana phases.
The second term of Eq.~(\ref{CD}), which is of  ${\cal O}(\alpha_e)$ with respect to the first,  depends on the neutrino mass scale and Majorana phases due to removing the  $\tau$ from  the loop below $m_\tau$, as for the penguin diagram.

The other 8  coefficients  in the Lagrangian of Eq.~(\ref{L1}) will be discussed  further in~\cite{pl2}.
The  coefficient on Gold,  $C_{Aheavy,L}$,    should be predicted  in  the type II seesaw, where    $\muc$ rates   are related to the $n/p$ ratio.
The remaining coefficients 
 are suppressed: for instance the dipole   $C^{e\mu}_{D,L}$  should be $\approx \frac{m_e}{m_\mu} C^{e\mu}_{D,R}$, as expected in  neutrino mass models where the  new particles only interact with  lepton doublets (the chirality-flip is via SM Yukawa interactions).  Similarly,  the  operators  with flavour-change  involving singlet leptons ($C_{S,RR}, C_{S,LL}, C_{V,RL},C_{V,RR},
 C_{Alight,R},C_{Aheavy,R}$) are not discussed here, because they are  Yukawa suppressed.
 So we already see that the type II seesaw predicts that more than half the coefficients of Eq.~(\ref{L1}) are negligible; however, many of  these predictions  are generic to  models  where the New Particles interact only with lepton doublets.

The type II seesaw is expected to     predict additional relations between  the Wilson  coefficients of Eq.~(\ref{L1}), because the flavour structure of  LFV is controlled by the neutrino mass matrix. This should allow to predict ratios of coefficients, despite that  the  overall magnitude of LFV 
is  unknown.
We focus on the remaining three coefficients, given in Eqs.~(\ref{CVLL}), (\ref{ping}) and~(\ref{CD}). These formulae  suggest the model prefers  a hierarchy $10^{-3}: 10^{-2}: 1$ between the dipole, penguin-induced  and tree-level coefficients; however, we aim to identify regions of coefficient space that  the model cannot predict, not what it prefers.

We observe that the tree-level  four-lepton coefficient $C_{V, LL}^{e\mu ee}$ given in Eq.~(\ref{CVLL}) can vanish,  either for
$[m_\nu]_{ee} \to 0$ in NH for  $m_{min} \sim \Delta_{sol}$
(as is familiar from  neutrinoless double $\b$-decay),
or  for  $[m_\nu]_{e\mu} \to 0$, which can occur for   any $m_{min} \gsim \Delta_{sol}$    in NH and IH  by suitable choice of both Majorana phases.
If  $C_{VLL}^{e\mu ee}$ vanishes,
 the dipole to penguin  ratio is predicted:
\bea
\frac{C^{e\mu}_{D, R}}{C^{e\mu ee}_{V, LR}} \approx
\frac{3e}{ 32 \pi\alpha_e  \ln \frac{M_\Delta}{m_\tau}}
\sim  \frac{2}{\pi^2} \, 
~~~.
\label{t2pred1}
\eea
When $[m_\nu]_{\mu e} \to 0$, this occurs because
 the Majorana phase and neutrino mass scale dependent terms of  the penguin and dipole  are proportional to $|[m_\nu]_{\mu e}|$ -- see the second terms of Eqs.~(\ref{ping}) and~(\ref{CD}).
  When
  $C_{VLL}^{e\mu ee}$  vanishes with   $[m_\nu]_{ee}$,
  this prediction  is also  approximately obtained:
  $\frac{C^{e\mu}_{D, R}}{C^{e\mu ee}_{V, LR}} \approx
  \frac{2}{\pi^2}\times [.66 \to 2]$,
  because    the second term of the penguin coefficient (which depends on Majorana phases and  the mass scale,
see Eq.~\ref{ping}) is  $\lsim 1/2 $ of  the first term, whereas the dipole is numerically unaffected.

It is also the case that  the ``penguin-induced'' coefficient of Eq.~(\ref{ping}), as well as the dipole coefficient Eq.~(\ref{CD}), can separately vanish for specific choices of  both  Majorana phases  and  the   neutrino mass scale
in the appropriate range (However, a high neutrino mass scale $m_{min} \gsim .2$ eV is required  for  the dipole coefficient to vanish.).  So in   all  limits where  one of the three coefficients  $C_{V, LL}^{e\mu ee}$, $C^{e\mu ee}_{V, LR}$ or $C^{e\mu}_{D, R}$ vanishes,   the ratio of the non-vanishing  coefficients is constrained\footnote{The values of the neutrino parameters that lead to cancellations in the coefficients are sensitive to the triplet mass. Therefore, the predictions/expectations for the non-vanishing coefficients ratio may significantly depend on the assumption $M_\Delta\sim 1$ TeV.}.  However,  the large coefficient ratios that arise when the dipole or  penguin vanishes  may be beyond the  sensitivity of upcoming experiments.

\begin{figure}[ht]
  \begin{center}
    \includegraphics[height=6.3cm,width=9cm]{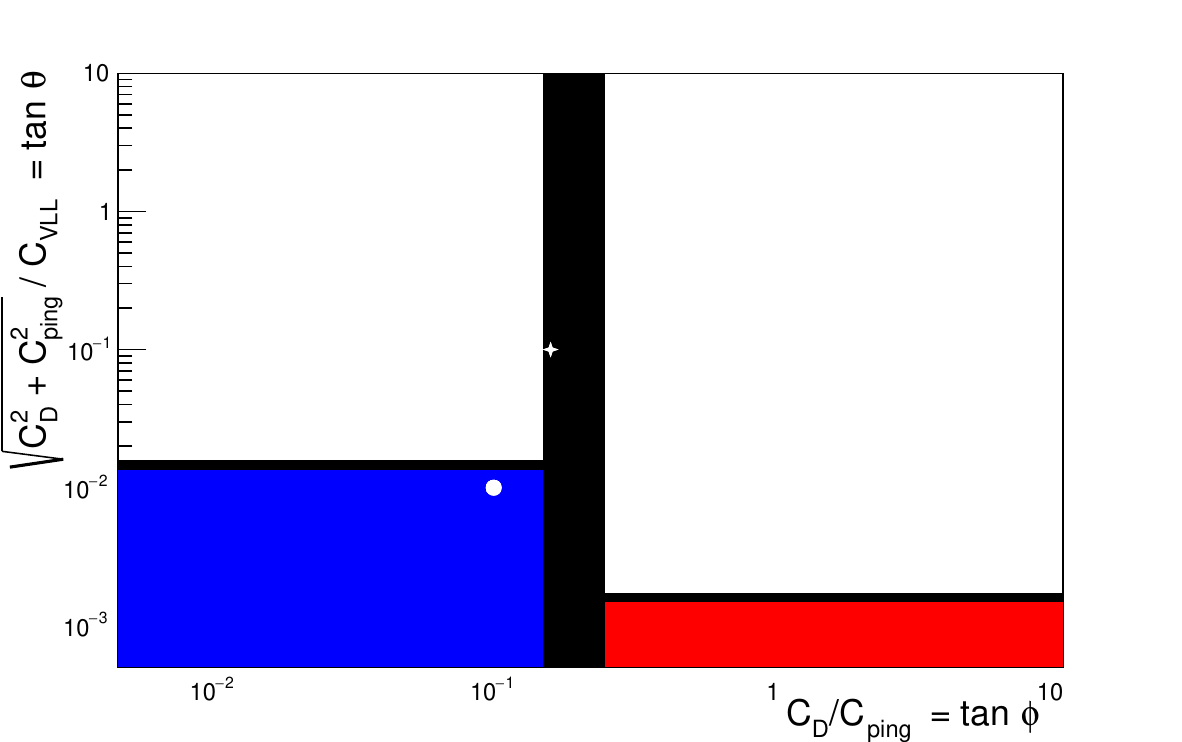}
  \end{center}
\caption{ The white regions indicate
  ratios  of operator coefficients  that  the type II seesaw {\it cannot} predict, as discussed after Eq.~(\ref{t2angles}),   where $\tan \theta$ and $\tan \phi$ are defined.
  Upcoming experiments are sensitive to the plotted  ranges of the ratios.   These estimates are independent of the neutrino mass hierarchy and mass scale; the star  and  circle  are located  respectively in  the regions predicted in NH and IH for $m_{min} = 0$  and specific choices of the Majorana phase
(but the regions can be larger when the phase varies; the IH region has filaments).
  \label{fig:angulart2} }
\end{figure}

  In order to graphically represent the  area of coefficient space that the
  type II seesaw model {\it cannot}  reproduce, we plot the magnitudes
  $|C^{e\mu}_{D, R}|$, $|C^{e\mu ee}_{V, LR}|$ and $|C^{e\mu ee}_{V, LL}|$ in spherical coordinates, with  on the $\hat{z}$ axis  $|C^{e\mu ee}_{V, LL}|\propto \cos \theta $.
  The current bounds  and the reach of upcoming experiments are given in table \ref{tab:Cbds}, which imply that upcoming experiments could probe 
\bea
\tan \theta \equiv \frac{\sqrt{|C^{e\mu}_{D, R}|^2 + |C^{e\mu ee}_{V, LR}|^2}}{|C^{e\mu ee}_{V, LL}|}  : 10^{-3} \to 10 ~~~,~~~
\tan \phi \equiv \frac{|C^{e\mu}_{D, R}|}{ |C^{e\mu ee}_{V, LR}|} : 10^{-2} \to 10
\label{t2angles}
\eea
Fig.~\ref{fig:angulart2} illustrates  (as empty) the  regions of the tree/penguin/dipole coefficient  space that are inaccessible to the type II seesaw model. 
The vertical bar represents the correlation between the dipole and penguin when the tree contribution  shrinks, given  in   Eq.~(\ref{t2pred1}).
For large tree contribution, the penguin contribution can shrink when the second term of Eq.~(\ref{ping}), $\propto |[m_\nu]_{\mu e}|$, cancels the first. This happens for values of the unconstrained neutrino parameters (the lightest neutrino mass and the Majorana phases) that enhance the tree-level coefficient $|C^{e\mu ee}_{V, LL}|$, so that $\tan \theta \lsim 10^{-3}$ -- this gives the upper bound to the red region.
Finally, for  generic values of the  Majorana phases, the  tree coefficient  is large with respect to the penguin-induced coefficients and the dipole, which  corresponds to the blue region at $\tan \theta\to 0$
and $\tan \phi  
\lsim 2/\pi^2$. In this paper, we   leave  the neutrino mass scale free, so  can obtain $\tan \phi 
\to 10^{-2}$ by increasing $m_{min}$ to $\gsim 0.2$ eV; we will study the impact of complementary observables -- such as the cosmological bound on the neutrino mass scale -- in~\cite{pl2}.

\section{Inverse  Type I seesaw }
\label{sec:t1}

In this section, we consider the inverse type I seesaw model~\cite{Wyler:1982dd,Mohapatra:1986aw,Mohapatra:1986bd}, which generates neutrino masses via the exchange of heavy  gauge-singlet fermions.  Like the type II seesaw, the model   can generate LFV without Lepton Number Violation,  so  LFV rates are  not suppressed by small neutrino masses. However, unlike the type II case, the flavour-changing couplings are disconnected from the neutrino mass matrix,  and several  heavy new particles are added, with potentially different masses.

We add to the SM $n$ pairs of gauge singlet fermions $N,S$ of opposite chirality, with the  interactions
\begin{equation}
	\delta\mathcal{L}_{NS}=i\overline{N}\slashed{\partial} N+i\overline{S}\slashed{\partial} S-\left(Y^{\alpha a}_\nu(\overline{\ell}_\alpha\tilde{H} N_a)+M_{ab}\overline{S}_a N_b+\frac{1}{2}\mu_{ab}\overline{S_a} S^c_b+{\rm h.c}\right),
\end{equation}
where $Y_\nu$ is a complex $3\times n$ dimensionless matrix and $M,\mu$ are $n\times n$ mass matrices. If Lepton Number is attributed to $\ell, N$ and $S$, then  only $\mu$ is Lepton Number Violating.  Upon the spontaneous breaking of the electroweak symmetry, the neutrino mass Lagrangian reads (suppressing flavour indices)
\begin{equation}
	\mathcal{M}_{\nu N}=\overline{\begin{pmatrix} \nu_L & N^c & S\end{pmatrix}}\begin{pmatrix}0 & m_D & 0 \\ 
		m_D^T & 0 & M^T \\
		0 & M & \mu \end{pmatrix} \begin{pmatrix} \nu^c_L \\
		N \\
		S^c \end{pmatrix}+{\rm h.c.}
\end{equation}
which, in the seesaw limit ($Y_\nu v= m_D\ll M$), give the following active neutrino masses
\begin{equation}
	m_\nu= m_D (M^{-1})\mu (M^T)^{-1} m^T_D.
\end{equation}
while for $M\gg \mu$ the $N,S$ pairs have pseudo-Dirac masses determined by the eigenvalues of $M$. Neutrino masses and oscillation parameters can be obtained by adjusting the lepton number breaking matrix $\mu$ for arbitrary choices of the Yukawa couplings $Y_\nu$ and sterile neutrino masses $M$.  
This contrasts with the ``vanilla'' type I seesaw expectation of GUT scale sterile neutrinos or suppressed Yukawa couplings~\cite{Minkowski:1977sc,Gell-Mann:1979vob,Yanagida:1979as,Glashow:1979nm,Mohapatra:1979ia}, which give negligible contributions to LFV observables. 

In the following, we consider $M\sim $ TeV and allow $Y_\nu$ to vary in the parameter space allowed by current LFV  searches and other experimental constraints. Low-scale type I seesaw models can be directly probed via the production of the {heavy} neutral leptons at colliders~\cite{delAguila:2007qnc, Atre:2009rg, Deppisch:2015qwa, Banerjee:2015gca, Antusch:2016vyf, Das:2018usr, Mekala:2022cmm}, or indirectly through the active-sterile neutrino mixing {(or the associated non-unitarity of the effective $3 \times 3$ lepton mixing matrix)}, which affect
{electroweak precision observables, universality ratios and lepton flavor violating processes}~\cite{Antusch:2014woa, Fernandez-Martinez:2016lgt, Chrzaszcz:2019inj, Blennow:2023mqx}.

\subsection{$\mu\to e$ LFV}
\begin{figure}[t]
	\centering
	\begin{subfigure}{0.6\textwidth}
		\begin{tikzpicture}[scale=1.2]
			\begin{feynman}
				\vertex (a) at (0, 2);
				\vertex (lin) at (-1, 2.25) {\(\mu_L\)};
				\vertex (b) at (1, 2);
				\vertex (Hlin) at (0.5, 2);
				\vertex (H) at (0.5, 2.75) {\(H\)};
				\vertex (H1) at (-0.15, 1.25)  {\(H\)};
				\vertex (gb1) at (0.5, 1.5);
				\vertex (gb2) at (0.5, 1);
				\vertex (f1) at (-1, 0.75) {\(f\)};
				\vertex (f2) at (2, 0.75) {\(f\)};
				\vertex (lout) at (2, 2.25) {\(e_L\)};
				
				\diagram* [small]{
					(lin) -- [fermion] (a) -- [fermion, edge label=\(N\)] (Hlin) --  [fermion] (b) -- [fermion]  (lout),
					(Hlin) -- [charged scalar] (H),
					(a) -- [anti charged scalar]  (gb1),
					(gb1) -- [photon] (b),
					(gb1) -- [anti charged scalar] (H1),
					(gb1) -- [photon, edge label=\(Z\)] (gb2), 
					(f1) -- [fermion] (gb2) -- [fermion] (f2),
				};
			\end{feynman}
		\end{tikzpicture}
		\begin{tikzpicture}[scale=1.2]
			\begin{feynman}
				\vertex (a) at (0, 2);
				\vertex (lin) at (-1, 2.25) {\(\mu_L\)};
				\vertex (b) at (1, 2);
				\vertex (Hline1) at (0.35, 2);
				\vertex (Hline2) at (0.65, 2);
				\vertex (gb1) at (0.5, 1.5);
				\vertex (gb2) at (0.5, 1);
				\vertex (f1) at (-1, 0.75) {\(f\)};
				\vertex (f2) at (2, 0.75) {\(f\)};
				\vertex (lout) at (2, 2.25) {\(e_L\)};
				
				\diagram* [small]{
					(lin) -- [fermion] (a) --[fermion, edge label=\(N\)] (b) -- [fermion] (lout),
					(a) -- [charged scalar, half right,  edge label=\(H^-\)]  (b),
					(gb1) -- [photon, edge label=\(\gamma\)]  (gb2), 
					(f1) -- [fermion] (gb2) -- [fermion] (f2),
				};
			\end{feynman}
		\end{tikzpicture}
		\caption{ $\mathcal{O}(Y_\nu Y^\dagger_\nu)$ $Z,\gamma$ penguin diagram contributing to $\mu\to e$ four-vector operators.}
		\label{fig:penguinYsq}
	\end{subfigure}\quad
	\begin{subfigure}{0.3\textwidth}
		\begin{tikzpicture}[scale=1.4]
			\begin{feynman}
				\vertex (a) at (0, 2);
				\vertex (lin) at (-1, 2.25) {\(\mu_L\)};
				\vertex (b) at (1, 2);
				\vertex (Hline1) at (0.35, 2);
				\vertex (H1) at (0.35, 2.75) {\(H\)};
				\vertex (Hline2) at (0.65, 2);
				\vertex (H2) at (0.65, 2.75) {\(H\)};
				\vertex (gb1) at (0.5, 1.5);
				\vertex (gb2) at (0.5, 1);
				\vertex (f1) at (-1, 0.75) {\(f\)};
				\vertex (f2) at (2, 0.75) {\(f\)};
				\vertex (lout) at (2, 2.25) {\(e_L\)};
				
				\diagram* [small]{
					(lin) -- [fermion] (a) --[fermion, edge label=\(N\)] (Hline1) -- [fermion] (Hline2) -- [fermion, edge label=\(N\)] (b) -- [fermion] (lout),
					(a) -- [anti charged scalar, half right]  (b),
					(gb1) -- [photon, edge label=\(Z\)]  (gb2), 
					(f1) -- [fermion] (gb2) -- [fermion] (f2),
					(Hline1) -- [charged scalar] (H1),
					(Hline2) -- [anti charged scalar] (H2),
				};
			\end{feynman}
		\end{tikzpicture}
		\caption{Penguin diagram proportional to four  neutrino Yukawas contributing to the $\mu\to e$ four-fermion operators}
		\label{fig:penguinfourY}
	\end{subfigure}
	\quad
	\begin{subfigure}{0.3\textwidth}
		\begin{tikzpicture}
			\begin{feynman}[large]
				\diagram* [inline=(c.base), horizontal=a to b] {
					i1 [particle=\(\mu_L\)]
					-- [fermion] a
					-- [anti charged scalar, edge label=\(H\)] b
					-- [fermion] f1 [particle=\(e_L\)],
					i2 [particle=\(e_L\)]
					-- [anti fermion] c
					-- [charged scalar, edge label'=\(H\)] d
					-- [anti fermion] f2 [particle=\(e_L\)],
					{ [same layer] a -- [fermion, edge label'=\(N\)] c },
					{ [same layer] b -- [anti fermion, edge label=\(N\)] d},
				};
			\end{feynman}
		\end{tikzpicture}
		\caption{Box diagrams matching onto the $\mu\to e_L \overline{e_L} e_L$ vector}
		\label{fig:boxll}
	\end{subfigure}
	\begin{subfigure}{0.3\textwidth}
		\begin{tikzpicture}
			\begin{feynman}[small]
				\vertex (mu) at (-1.5,0) {\(\mu_R\)};
				\vertex (e) at (3,0) {\(e_L\)};
				\vertex (H1) at (-0.5,1) {\(H\)};
				\vertex (H2) at (-0.5,0);
				\vertex (a)  at (0,0) ;
				\vertex (b) at (2,0) ;
				\vertex (boson1) at (1.5, 1);
				\vertex (boson2) at (2, 2);
				\vertex (c) at (1.3,1.1) ;
				\vertex (H3) at (2.2,1.6) {\(\gamma\)};
				\diagram* [inline=(a.base)]{
					(H1) -- [charged scalar] (H2),
					(H3) -- [photon] (c),
					(mu) -- [fermion] (a),
					(b) -- [fermion] (e),
					(a) -- [fermion, half left, edge label'=\(N\)] (b)  -- [scalar, edge label=\(H\)] (a),
				};
			\end{feynman}
		\end{tikzpicture}
		\caption{Matching onto the $\mu\to e$ dipole operator}\label{fig:dipolestypeI}
	\end{subfigure}
	\caption{Matching contributions to $\mu\to e$ operators in the inverse seesaw. The diagrams illustrate the relevant interactions that are generated, but other diagrams may also contribute to the same operators~\label{fig:diagramstypeI}}
\end{figure}
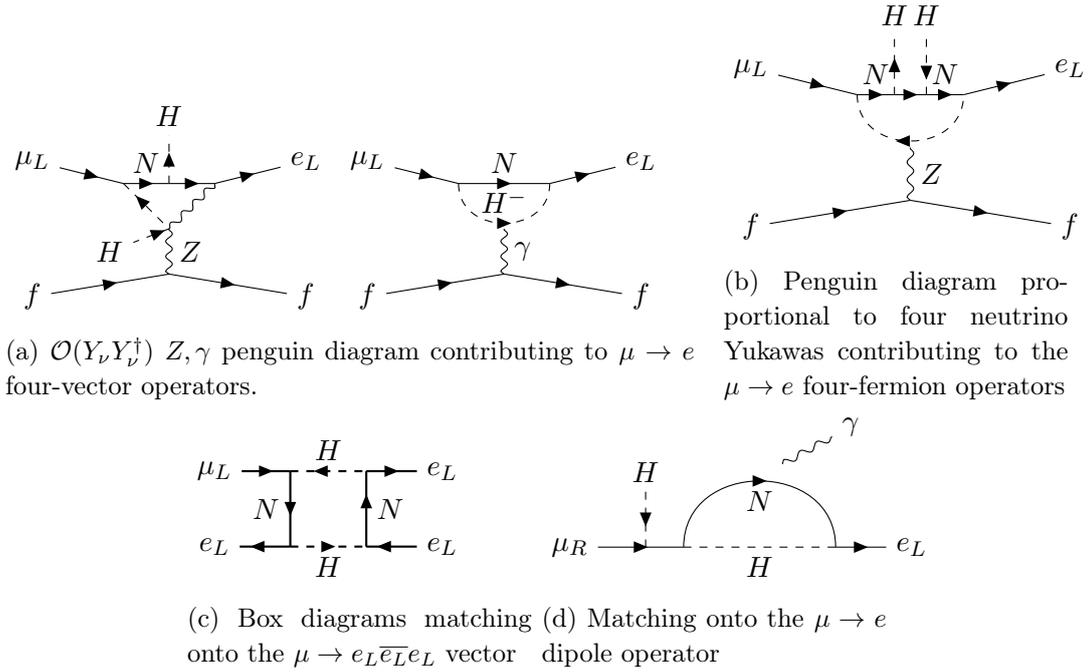
Large lepton flavour violating transitions are among the distinctive features of the inverse seesaw~\cite{Tommasini:1995ii,Ilakovac:1994kj,Ibarra:2011xn,Dinh:2012bp,Alonso:2012ji,Abada:2015oba,Coy:2018bxr,Abada:2021zcm,Zhang:2021jdf,Coy:2021hyr,Granelli:2022eru,Crivellin:2022cve}.  In this paper, we focus on the contact interactions that are relevant for $\mu\to e$ observables and aim at determining the region of the EFT coefficient space that the model cannot reach. 

The LFV transitions we are interested in occurs in this model via loops, as we illustrate in Fig.~\ref{fig:diagramstypeI}.  The four-fermion operator coefficients  are obtained in matching out the heavy singlets in penguin and box diagrams.
The  vector four-fermion coefficients   $C^{e\mu ff}_{V,LX}$,   receive contributions from penguin diagrams shown in Figs.~\ref{fig:penguinYsq} and~\ref{fig:penguinfourY},
which are respectively ${\cal O}(Y_\nu Y_\nu^\dagger)$ and ${\cal O}(Y_\nu Y_\nu^\dagger Y_\nu Y_\nu^\dagger)$.
We include the diagram in Fig.~\ref{fig:penguinfourY} following Ref.~\cite{Crivellin:2022cve}, who observed that the  contributions $\propto Y^4_\nu$ could be relevant for $Y_\nu$   $\mathcal{O}(1)$.  
The box diagrams of Fig. \ref{fig:boxll} also match onto  vector four-lepton operators, while the diagrams of Fig.~(\ref{fig:dipolestypeI}) match onto the $\mu\to e$ dipole.  Similarly to the type II seesaw model of Section~\ref{sec:t2}, the new states couple to the left-handed doublets, so  the  operators featuring LFV currents with  electron singlets are suppressed by the electron Yukawa coupling.
 As a result, the model matches onto five of the operators in Eq.~(\ref{L1}). Leaving aside the one associated with $\mu \to e$ conversion on heavy nuclei, since upcoming experiments will use light targets, we are left with the following four coefficients\footnote{ $C^{e\mu}_{Aheavy,L}$ can be predicted from these four coefficients, and will be given in~\cite{pl2}.}:
\bea
C^{e\mu ee}_{V,LR}&\simeq & v^2\frac{\alpha_e}{4\pi}\bigg(1.5[Y_\nu M_a^{-2}\left(\frac{11}{6}+\ln\left(\frac{m^2_W}{M_a^2}\right)\right)Y_\nu^\dagger]_{e\mu} -2.7 [Y_{\nu}(Y^\dagger_\nu Y_\nu)_{ab}\frac{1}{M^2_{a}-M^2_{b}}\ln\left(\frac{M^2_{a}}{M^2_{b}}\right)Y_\nu^\dagger]_{e\mu}\nonumber\\
&+&\mathcal{O}\left(\frac{\alpha_e}{4\pi}\right)\bigg)\nonumber\\
C^{e\mu}_{Alight,L}&\simeq & v^2\frac{\alpha_e}{4\pi}\bigg(-0.6[Y_\nu M_a^{-2}\left(\frac{11}{6}+\ln\left(\frac{m^2_W}{M_a^2}\right)\right)Y_\nu^\dagger]_{e\mu} +1.1 [Y_{\nu}(Y^\dagger_\nu Y_\nu)_{ab}\frac{1}{M^2_{a}-M^2_{b}}\ln\left(\frac{M^2_{a}}{M^2_{b}}\right)Y_\nu^\dagger]_{e\mu}\nonumber\\
&+&\mathcal{O}\left(\frac{\alpha_e}{4\pi}\right)\bigg)\nonumber\\
C^{e\mu ee}_{V,LL}&\simeq & v^2\frac{\alpha_e}{4\pi}\bigg(-1.8[Y_\nu M_a^{-2}\left(\frac{11}{6}+\ln\left(\frac{m^2_W}{M_a^2}\right)\right)Y_\nu^\dagger]_{e\mu} + 2.7 [Y_{\nu}(Y^\dagger_\nu Y_\nu)_{ab}\frac{1}{M^2_{a}-M^2_{b}}\ln\left(\frac{M^2_{a}}{M^2_{b}}\right)Y_\nu^\dagger]_{e\mu}
\nonumber\\
&+& 2.5 Y^{e a}_\nu Y^{*\mu a}_\nu Y^{e b}_\nu Y^{*e b}_\nu\frac{1}{M^2_{a}-M^2_{b}}\ln\left(\frac{M^2_{a}}{M^2_{b}}\right)+\mathcal{O}\left(\frac{\alpha_e}{4\pi}\right)\bigg)\nonumber \\
C^{e\mu}_{D,R}&\simeq &-\frac{v^2}{2}\left(\frac{\alpha_e}{4\pi e}\right)[Y_\nu M^{-2}Y_\nu^\dagger]_{e\mu},\label{eq:invseesawnondeg}
\eea
where $a,b$ are summed over the number $n$ of sterile neutrinos.
We include the finite part of the penguin diagrams shown in Fig.~\ref{fig:penguinYsq} because the ratios of sterile masses and the electroweak scale involved in the logarithms are not large (as we discuss in the introduction). Higher-order terms in the $\alpha_e/(4\pi)$ expansion are neglected because they are small and would require including some two-loop diagrams for a consistent treatment. Consequently, the results presented in Eq.~(\ref{eq:invseesawnondeg}) are reliable at the { $\lesssim 10\%$ level}.

The above  coefficients, generated by the model and in principle observable,  are linear combinations of four contractions of the Yukawa and sterile neutrino mass matrices { (which we refer to as ``invariants'')}. Since the number of coefficients equals the number of invariants, it seems  that the model could predict any observation -- i.e. any point in the 4-dimensional space of the operator coefficients -- with suitable choices of the $Y_\nu$ and $M$ matrices. { However,} the number of invariants  { is} reduced if the sterile neutrinos are nearly degenerate~\footnote{A motivation for considering this limit comes from the baryon asymmetry of the Universe, which can be generated from resonant leptogenesis with highly degenerate TeV-scale sterile neutrinos (see e.g. Refs.~\cite{Pilaftsis:2003gt,Blanchet:2009kk,BhupalDev:2014pfm,daSilva:2022mrx}), or from the CP-violating oscillations of nearly degenerate sterile neutrinos~\cite{Akhmedov:1998qx} with masses in the GeV~\cite{Asaka:2005pn} to multi-TeV~\cite{Klaric:2021cpi} range.}. In this limit, the combination entering the $\mathcal{O}(Y_\nu Y^\dagger_\nu)$ penguin contributions aligns with the matrix elements parameterizing the dipole coefficient. Indeed, by expanding $M^2_a/M^2=1+x_a$ for small $x_a$  (where $M$ now denotes the average sterile neutrino mass), we have that
\begin{align}
	\frac{1}{M_a^2}\left(\frac{11}{6}+\ln\left(\frac{m_W^2}{M^2_a}\right)\right)&=\frac{1}{M^2(1+x_a)}\left(\frac{11}{6}+\ln\left(\frac{m_W^2}{M^2}\right)-\ln(1+x_a)\right)\nonumber\\
	&=\frac{1}{M^2}\left(\frac{11}{6}+\ln\left(\frac{m_W^2}{M^2}\right)+\mathcal{O}(x_a)\right).\label{degen1}
\end{align} 
If the mass-splitting between the heavy singlets is $\lesssim v^2$, the error introduced by the degenerate approximation is a dimension eight $v^2/M^2$ suppressed contribution, that, for TeV scale sterile masses, would be approximately of the same order of the neglected $\mathcal{O}(\alpha_e/4\pi)$ corrections. Similarly, the leading order term in the $x_a$ expansion of the mass function that enters in the $\mathcal{O}(Y_\nu Y_\nu^\dagger Y_\nu Y_\nu^\dagger)$ penguin and in the boxes is
\begin{equation}
	\frac{1}{M^2_{a}-M^2_{b}}\ln\left(\frac{M^2_{a}}{M^2_{b}}\right)=\frac{1}{M^2}\left(1+\mathcal{O}(x_a, x_b)\right),
	\label{degen2}
\end{equation}
so that in the nearly degenerate limit, we find\footnote{Recall that the operator coefficients depend logarithmically on the scale of the new states, which we take to be around $1\, {\rm TeV}$. } 
\begin{align}
	C^{e\mu}_{D,R}(m_\mu)&\simeq -10^{-3} \frac{v^2}{M^2}(Y_\nu Y^\dagger_\nu)_{e\mu}\nonumber \\
	C^{e\mu}_{Alight,L}(m_\mu)&\simeq\frac{v^2}{M^2}\left( 10^{-3} (Y_\nu Y^\dagger_\nu)_{e\mu}+6.6\times 10^{-4}(Y_\nu Y^\dagger_\nu Y_\nu Y^\dagger_\nu)_{e\mu}\right) \nonumber \\
	C^{e\mu ee}_{V,LR}(m_\mu)&\simeq\frac{v^2}{M^2}\left(-2.8\times 10^{-3}(Y_\nu Y^\dagger_\nu)_{e\mu}-1.6\times 10^{-3}(Y_\nu Y^\dagger_\nu Y_\nu Y^\dagger_\nu)_{e\mu}\right)\nonumber \\
	C^{e\mu ee}_{V,LL}(m_\mu)&\simeq\frac{v^2}{M^2}\left(3.3\times 10^{-3}(Y_\nu Y^\dagger_\nu)_{e\mu}(1+0.56 (Y_\nu Y^\dagger_\nu)_{ee})+1.55\times 10^{-3}(Y_\nu Y^\dagger_\nu Y_\nu Y^\dagger_\nu)_{e\mu}\right) \label{eq:invseesawdeg}
\end{align}
Despite the large number of free parameters in the inverse seesaw model, even in the degenerate limit, the coefficients of the $\mu\to e$ operators can now be determined by just two invariant contractions of the neutrino Yukawa matrix. Being linear combinations of two invariants, the correlations of the operator coefficients that the model can predict are restricted: by measuring two (complex) coefficients, it would be possible to predict the others. 
Focusing on the first three operators of Eq.~(\ref{eq:invseesawdeg}), we find that 
\begin{figure}[ht]
	\centering
	\begin{subfigure}{.4\textwidth}
		\centering
		\includegraphics[width=\linewidth]{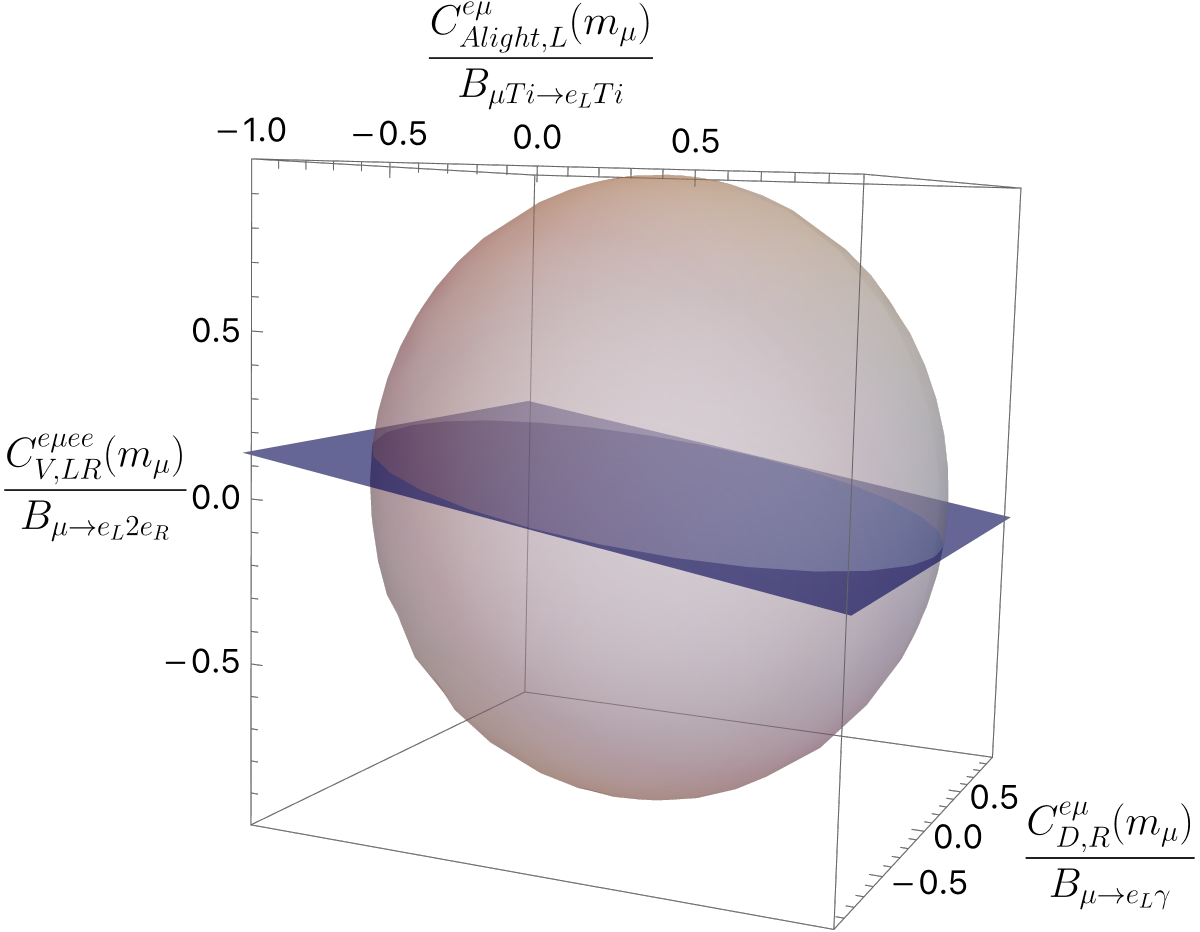}
		\caption{The light-gray sphere illustrate the experimentally allowed ellipse in the $C^{e\mu}_{D,R}(m_\mu),C^{e\mu e e}_{Alight,L}(m_\mu),C^{e\mu e e}_{V,LR}(m_\mu)$  space. We consider the real parts of the coefficients and normalise to the current upper bound. If the sterile neutrinos are nearly degenerate, the model can cover the region defined in Eq.~(\ref{eq:typeIcvlr}), which correspond to the blue plane.}\label{fig:cvlr}
	\end{subfigure}\qquad
	\begin{subfigure}{.4\textwidth}
		\centering
		\includegraphics[width=\linewidth]{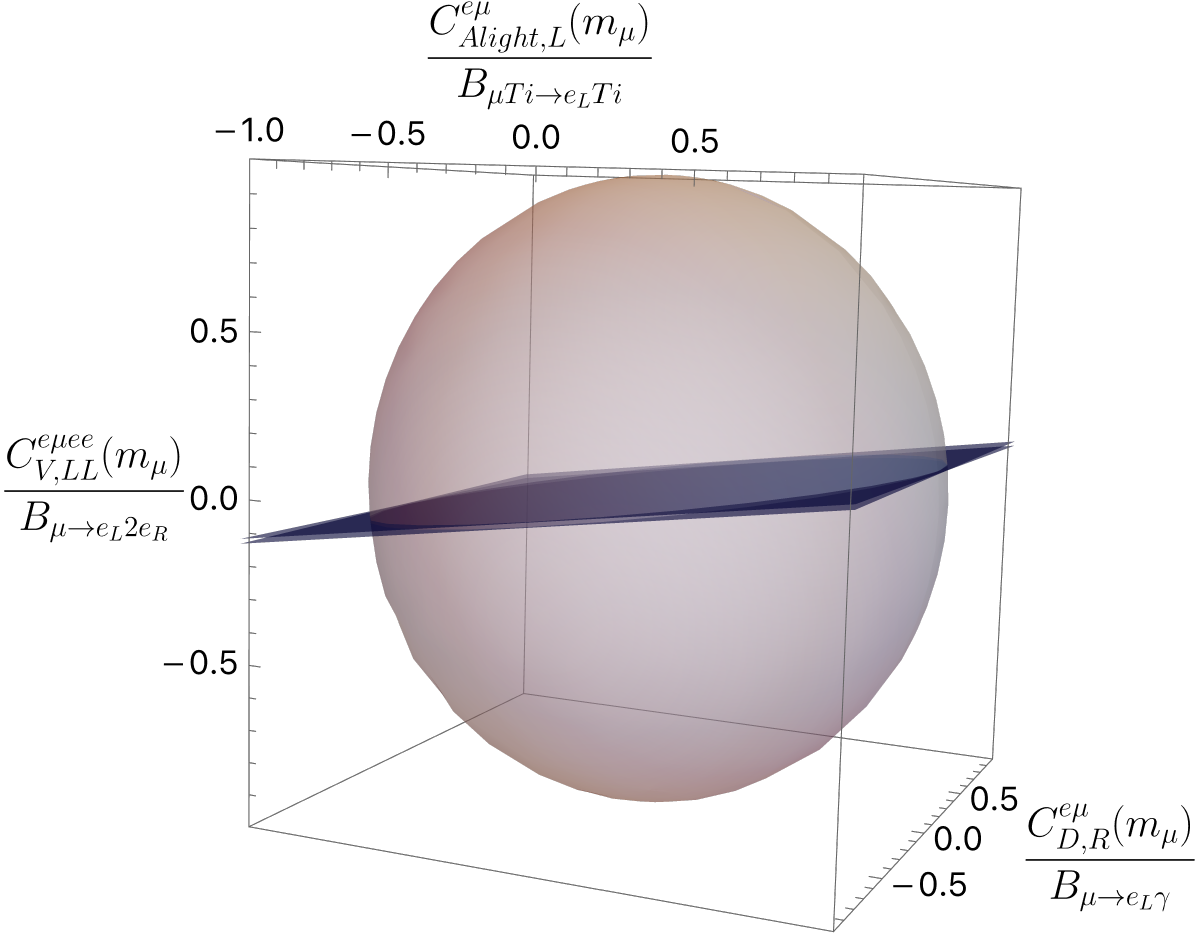}
		\caption{The light-gray sphere illustrate the experimentally allowed ellipse in the $C^{e\mu}_{D,R}(m_\mu),C^{e\mu e e}_{Alight,L}(m_\mu),C^{e\mu e e}_{V,LL}(m_\mu)$  space. We consider the real parts of the coefficients and normalise to the current upper bound. If the sterile neutrinos are nearly degenerate, the model can cover the region defined in Eq.~(\ref{eq:typeIcvll}) and correspond to the volume delimited by the two blue planes (see Fig.~\ref{fig:cvllzoom})}\label{fig:cvll}
	\end{subfigure}\\
	\begin{subfigure}{.4\textwidth}
		\centering
		\includegraphics[width=\linewidth]{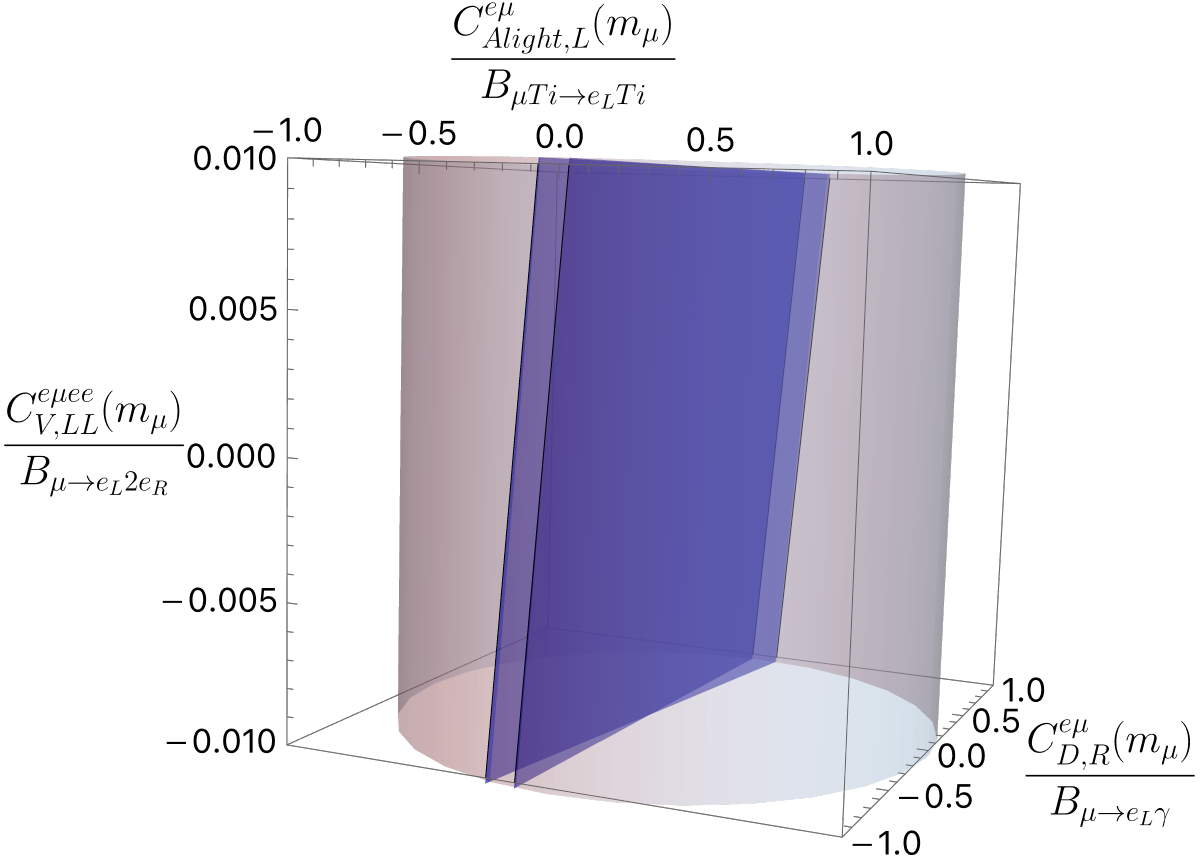}
		\caption{Zoom of Fig.~\ref{fig:cvll}. The model can cover only the region enclosed by the two planes.}\label{fig:cvllzoom}
	\end{subfigure}
	\caption{Parameter space covered by the inverse seesaw (with degenerate sterile neutrinos) in the low-energy operator coefficient space}
	\label{fig:typeIcv}
\end{figure}
\begin{equation}
	C^{e\mu ee}_{V,LR}(m_\mu)=-2.4 C^{e\mu}_{Alight,L}(m_\mu)+0.02C^{e\mu}_{D,R}(m_\mu)\label{eq:typeIcvlr}
\end{equation}
In the purely left-handed $\mu\to 3e$ vector the magnitude of the coefficient multiplying the matrix element $(Y_\nu Y^\dagger_\nu)_{e\mu}$ is dependent on the real and positive parameter $(Y_{\nu}Y^\dagger_\nu)_{ee}$ arising from the box diagram contribution. However, since the Yukawa couplings are assumed to be perturbative 
$(Y_{\nu}Y^\dagger_\nu)_{ee}\lesssim 1$, we can similarly find that
\begin{equation}
	C^{e\mu ee}_{V,LL}(m_\mu)=2.4 C^{e\mu}_{Alight,L}(m_\mu)+c_dC^{e\mu}_{D,R}(m_\mu)\label{eq:typeIcvll}
\end{equation}
where $-1.99\lesssim c_d\lesssim-0.57$. The correlations described by Eqs.~(\ref{eq:typeIcvlr}) and~(\ref{eq:typeIcvll}) hold, within the accuracy of our calculations, for general complex coefficients.  To visually represent the parameter space accessible to the inverse seesaw model, we consider the real parts of the coefficients and plot the corresponding planes in the 3D space of low-energy coefficients. By normalizing each coefficient to the upper limit imposed by current experimental searches, the allowed region of parameter space correspond to the interior of a sphere. The inverse seesaw model (with nearly degenerate sterile neutrinos) can sit in the intersection of this region with the planes defined by Eq.~(\ref{eq:typeIcvlr}) and Eq.~(\ref{eq:typeIcvll}), as illustrated in Fig.~\ref{fig:typeIcv}. Since the dipole coefficient in Eq.~(\ref{eq:typeIcvll}) is unknown but bounded, the model can cover the volume enclosed by the two extreme planes



\section{Leptoquark }
\label{sec:lq}

  This section studies the $\me$   predictions of an SU(2) singlet leptoquark
 of hypercharge $Y= 1/3$ that could fit the  the R$_{D}$ anomaly~\cite{BaBar:2012obs,Belle:2015qfa,Belle:2019rba,LHCb:2023zxo,LHCb:2023cjr,Sakaki:2013bfa,Cai:2017wry,Angelescu:2018tyl,Lee:2021jdr}, which is an  excess of $b\to c\bar{\tau}\nu$  events. Requiring the leptoquark to fit $R_D$ fixes the mass  to be {\cal O}(TeV) and restricts the quantum numbers,
 but  our   $\me$ interactions  are independent of the couplings that contribute to $R_D$.
Unlike the  models of the previous sections,  the leptoquark  couples to both lepton doublets and singlets, and   can mediate $\muc$ at tree level -- but does not generate neutrino masses.

The SU(2)-singlet leptoquark is denoted   $S_1$~\cite{Buchmuller:1986zs} (not to be confused with the singlet fermions  $\{S_a\}$ of the previous section), with interactions: 
\bea
{\cal L}_S & = & 
(D_\r S_1)^\dagger D^\r S_1 - m_{LQ}^2 S_1^\dagger S_1  + 
(- \lambda^{\a j}_{L} \overline{\ell}_\a  i \tau_2 q_j^c
+ \lambda^{\a j}_{R} \overline{e}_\a u_j^c ) S_1 
+  (\lambda^{\a j*}_{L} \overline{q^c}_j  i \tau_2  \ell_\a +
\lambda^{ \a j*}_{R} \overline{u^c}_j e_\a) S_1^{\dagger} \nonumber\\
&&+{ \rm ~Higgs~interactions}
\nonumber
\eea
where the leptoquark mass is  $m_{LQ}\simeq$ TeV, 
the  generation indices are $\a \in\{ e,\mu,\tau \}$
and $j\in \{u,c,t\}$, and the
sign of the doublet  contraction
is taken to give $+ \lambda^{\alpha j}_L \overline{e_L} (u_L)^c S_1$. {Like in the type II model of Section~\ref{sec:t2}, the leptoquark-Higgs interactions are neglected because their contributions to LFV observables are negligible assuming perturbative couplings.}

  Leptoquarks are strongly interacting, so can be readily produced at hadron colliders; the current  LHC searches impose $m_{LQ} \gsim$ 1-2 TeV~\cite{ParticleDataGroup:2022pth}.
  Also, their peculiar Yukawa interactions connecting quarks to leptons,  can predict   diverse quark and/or lepton  flavour-changing  processes~\cite{Crivellin:2020mjs,Lee:2021jdr,Coy:2021hyr}.
  For instance, non-zero $ \lambda^{\mu u}_{X}$,
 $ \lambda^{\mu c}_{X}, \lambda^{e u}_{X}$ and $\lambda^{e c}_{X}$
  induce   $\mu \to e$ processes on a $u$  and $c$ quark currents -- which we study here -- and  also induce  LFV $D$  decays  with
  $e^\pm \mu^\mp$ in the final state. 
 In addition,  $S_1$ will    mediate $\Delta F = 2$ four-quark operators via box diagrams which  can  contribute to meson-anti-meson mixing~\cite{UTfit:2007eik}.
We did not find relevant constraints on the  LFV interactions of $S_1$ from quark flavour physics,  but will discuss in more detail the complementarity of quark and lepton observables in~\cite{pl2}.

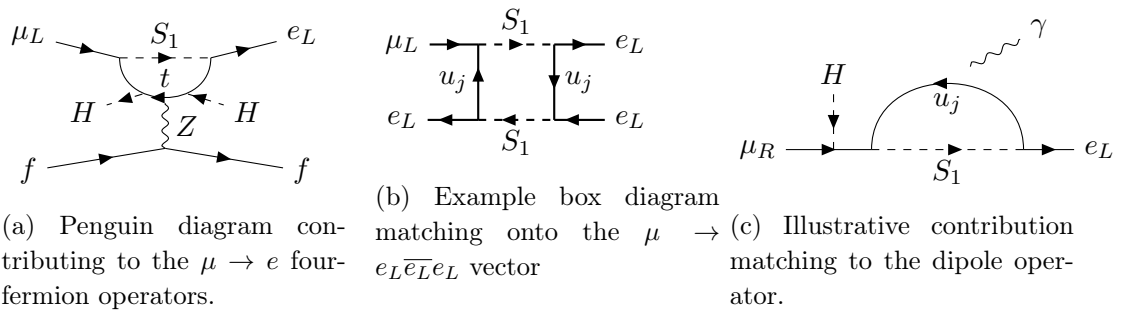
\begin{figure}[htb]
   \centering
   	\begin{subfigure}{0.3\textwidth}
 		\begin{tikzpicture}[scale=1.2]
 			\begin{feynman}
 				\vertex (a) at (0, 2);
 				\vertex (lin) at (-1, 2.25) {\(\mu_L\)};
 				\vertex (b) at (1, 2);
 				\vertex (Hline1) at (0.25, 1.6);
 				\vertex (H1) at (-0.4, 1.4) {\(H\)};
 				\vertex (Hline2) at (0.75, 1.6);
 				\vertex (H2) at (1.4, 1.4) {\(H\)};
 				\vertex (gb1) at (0.5, 1.5);
 				\vertex (gb2) at (0.5, 1);
 				\vertex (f1) at (-1, 0.75) {\(f\)};
 				\vertex (f2) at (2, 0.75) {\(f\)};.
 				\vertex (lout) at (2, 2.25) {\(e_L\)};
 				
 				\diagram* [small]{
 					(lin) -- [fermion] (a) -- [charged scalar, edge label=\(S_1\)] (b) -- [fermion] (lout),
 					(a) -- [anti fermion, half right,  edge label=\(t\)]  (b),
 					(gb1) -- [photon, edge label= \(Z\)] (gb2), 
 					(f1) -- [fermion] (gb2) -- [fermion] (f2),
 					(Hline1) -- [ charged scalar] (H1),
 					(Hline2) -- [anti charged scalar] (H2),
 				};
 			\end{feynman}
 		\end{tikzpicture}
 		\caption{Penguin diagram contributing to the $\mu\to e$ four-fermion operators.}
 		\label{fig:penguinlq}
 	\end{subfigure}\quad
 	\begin{subfigure}{0.3\textwidth}
 		\begin{tikzpicture}
 			\begin{feynman}[large]
 				\diagram* [inline=(c.base), horizontal=a to b] {
 					i1 [particle=\(\mu_L\)]
 					-- [fermion] a
 					-- [charged scalar, edge label=\(S_1\)] b
 					-- [fermion] f1 [particle=\(e_L\)],
 					i2 [particle=\(e_L\)]
 					-- [anti fermion] c
 					-- [anti charged scalar, edge label'=\(S_1\)] d
 					-- [anti fermion] f2 [particle=\(e_L\)],
 					{ [same layer] a -- [anti fermion, edge label'=\(u_j\)] c },
 					{ [same layer] b -- [fermion, edge label=\(u_j\)] d},
 				};
 			\end{feynman}
 		\end{tikzpicture}
 		\caption{Example box diagram matching onto the $\mu\to e_L \overline{e_L} e_L$ vector}
 		\label{fig:boxlq}
. 	\end{subfigure}
 	\begin{subfigure}{0.3\textwidth}
 		\begin{tikzpicture}
 			\begin{feynman}[small]
 				\vertex (mu) at (-1.5,0) {\(\mu_R\)};
 				\vertex (e) at (3,0) {\(e_L\)};
 				\vertex (H1) at (-0.5,1) {\(H\)};
 				\vertex (H2) at (-0.5,0);
 				\vertex (a)  at (0,0) ;
 				\vertex (b) at (2,0) ;
 				\vertex (boson1) at (1.5, 1);
 				\vertex (boson2) at (2, 2);
 				\vertex (c) at (1.3,1.1) ;
 				\vertex (H3) at (2.2,1.6) {\(\gamma\)};
 				\diagram* [inline=(a.base)]{
 					(H1) -- [charged scalar] (H2),
 					(H3) -- [photon] (c),
 					(mu) -- [fermion] (a),
 					(b) -- [fermion] (e),
 					(a) -- [anti fermion, half left, edge label'=\(u_j\)] (b)  -- [anti charged scalar, edge label=\(S_1\)] (a),
 				};
 			\end{feynman}
 		\end{tikzpicture}
 		\caption{Illustrative  contribution matching  to the dipole operator.}\label{fig:dipolelq}
 	\end{subfigure}
 	\caption{Representative diagrams for the matching  of the leptoquark onto four-fermion operators, and the dipole.~\label{fig:LQdiagrams}}
 \end{figure}

\begin{figure}[ht]
    \vspace{-.3cm}
    \includegraphics[scale=0.9]{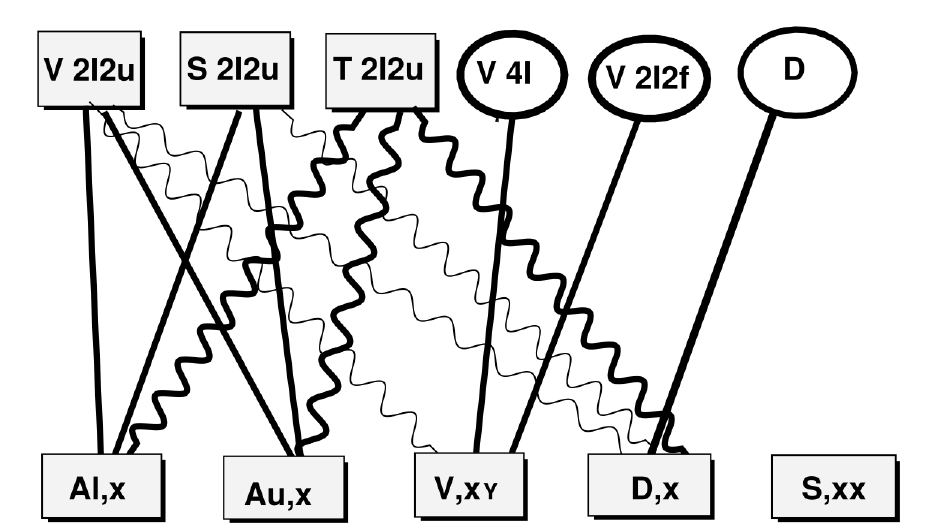}
 \caption{A schematic representation of how the leptoquark generates $\meg$, $\meee$ and $\muc$. The top row represents {\it  classes} of coefficients, generated in matching out the leptoquark,  of given  Lorentz structure and  particle type, for any  flavours and chiralities. 
The boxes  correspond to operators with coefficients  of ${\cal O}(\lambda^2/m_{LQ}^2)$, whereas the ovals  have  suppressed coefficients   $\sim {\cal O}(\lambda^2/[16\pi^2 m_{LQ}^2])$, or $\sim {\cal O}(\lambda^4/[16\pi^2 m_{LQ}^2])$.
The bottom row of boxes are the six observable coefficients (for fixed $e$ chirality  in the $\me$ bilinear) of Eq.~(\ref{L1}).
Lines represent the  transformation between the $m_{LQ}$ and  experimental scale; a straight line  means  the observable coefficient  can be  directly obtained in matching. Operator mixing  is represented as  wavy lines: a thick line
indicates    an ${\cal O}(1)$  contribution of at least one  operator from the class to the observable; a thin line  indicates a  more suppressed ${\cal O}(\a)$ contribution.
\label{fig:RGEs} }
\end{figure}

  In matching the leptoquark onto the QCD$\times$QED-invariant EFT at $m_{LQ}$,
vector 
  ($\propto \lambda^{*}_{R}\lambda_{R}$, $ \lambda^{*}_{L}\lambda_{L}$),
and scalar/tensor
($\propto \lambda^{*}_{R}\lambda_{L}$, $ \lambda^{*}_{L}\lambda_{R}$)
operators are generated at tree-level. 
We only consider the subset  which are quark flavour-diagonal and  $\me$ flavour-changing.
The model   matches onto  vector four-fermion operators of the form
$(\bar{e}\g^\rho P_X \mu) (\bar{f} \g_\rho P_Y f)$  (where $X,Y \in \{L,R\}$ and $f$ any  lepton or quark) via ``penguin'' diagrams (see Fig.~\ref{fig:penguinlq}), 
 and   also can generate  vector four lepton operators via box diagrams as in  Fig.~\ref{fig:boxlq}. Finally, the dipole operators can be generated  via the last diagram of Fig.~\ref{fig:LQdiagrams}. This collection of operators at the leptoquark mass scale
 is schematically represented in Fig.~\ref{fig:RGEs} as the top row of boxes and  ovals.

 Several of the operators generated in matching  out the leptoquark  are present in the Lagrangian of Eq.~(\ref{L1}).  For instance,  $S_1$ matches onto vector and/or scalar $\bar{e}$-$\mu$-$\bar{u}$-$u$ operators, which
 give large contributions to $\muc$.
 In addition,  the log-enhanced loops
 change the predictions  significantly:  the coefficients of  scalar  and tensor quark operators   respectively grow and shrink due to QCD,
 and  QED loops  can  cause some ${\cal O}(1)$ mixing, such as   the top and charm tensors into the dipole, or the $u$-tensor into the $u$-scalar.  The effect of the RGEs is represented  by lines in Fig.~\ref{fig:RGEs}.

At the experimental scale, the $S_1$ leptoquark generates both  $\mu \to e_L$ and $\mu \to e_R$ coefficients. For conciseness, we give results for  $\mu \to e_L$; the  $\mu \to e_R$ coefficients can be  obtained by judiciously interchanging $R\leftrightarrow L$.  The contribution to the dipole coefficients is
\bea
 \frac{ m_{LQ}^2}{ v^2}C^{e\mu}_{D,R} (m_\mu) &\simeq& 
  \frac{ e  [\lambda_L\lambda^\dagger_L]^{e\mu} }{128\pi^2}
\left(1- 16 \frac{\alpha_e}{4\pi} \ln\frac{m_{LQ}}{m_\mu}\right)
+ \frac{ 2 \alpha^2_e }{9 \pi^2 e} 
\left[ \lambda_L \ln\frac{m_{LQ}}{m_Q}  \lambda^\dagger_L \right]^{e\mu}  
 \nonumber \\&&
 -\frac{ \alpha_e  }{2\pi e y_\mu}
 \left[ \lambda_L Y_u  \tilde{f}^Q \ln\frac{m_{LQ}}{m_Q} \lambda^\dagger_R \right]^{e\mu} 
 ~~~~~~~~~~~~~
 \label{megLQ}
 \eea
where the first term is the matching contribution (times  its QED running),
the second term is the 2-loop  mixing of tree vector operators into the dipole,
the last term is the RG-mixing of   tensor operators to dipoles, and  the $m_Q$  serving as lower cutoff for the logarithms  (here and further in the paper) is max$\{ m_Q, 2~{\rm GeV}\}$ \footnote{We neglect  the estimates of
Ref.~\cite{Dekens:2018pbu}.}. For $C^{e\mu}_{D,L}$, one interchanges $ R\leftrightarrow L$.   
The QCD running  of the  quark tensor operator is intricated with the
QED mixing to the dipole~\cite{Bellucci:1981bs,Buchalla:1989we}, so induces a quark-flavour-dependent rescaling $\tilde{f}^Q \simeq \{1, 1.4\}$ for $\{t,c\}$ quarks.

  The leptoquark also  generates vector  four-lepton operators (for $X\in \{L,R\}$)
\bea
 \frac{ m_{LQ}^2}{ v^2} C_{V,LX}^{e\mu ee} (m_\mu) & \simeq &
-\frac{N_c }{64\pi^2} 
[\lambda_L \lambda_L^\dagger]^{e \mu} [\lambda_X \lambda_X^\dagger]^{ee}
     \left( 1\mp 12 \frac{\a_e}{4\pi} \ln\frac{m_{LQ}}{m_\mu} \right) 
{ +} \frac{\a_e}{3\pi} \left[ \lambda_L  \ln\frac{m_{LQ}}{m_Q}   \lambda_L^\dagger \right]^{e\mu}
\nonumber\\
&&
- g^e_X \frac{N_c }{16\pi^2} 
 \left[ \lambda_L Y_u \ln \frac{m_{LQ}}{m_Q} Y_u^\dagger   \lambda^\dagger_L \right]^{e\mu}
 \label{CVLLeeLQ}
 \eea
where  $g^e_L= -1+2\sin^2\theta_W$, $g^e_R=2\sin ^2\theta_W$,
the first  term represents   the box  diagram at $m_{LQ}$  (and its QED running to $m_\mu$, with $-$/$+$ for X=/$\neq$Y) which is represented as the $V,4l$ oval  at the top  of  Fig.~\ref{fig:RGEs} connecting to the $V_{XY}$ box at the bottom,
the second term is 
the  log-enhanced  photon penguin  that mixes  the 
tree operators  ${\cal O}_{VLL}^{QQ}$ (for $Q\in \{u,c,t\}$)
{into 4-lepton operators (represented in Fig.~\ref{fig:RGEs} as a thin wavy line between  the $V,2l2u$ and $V_{XY}$ boxes), and the last term   is  the contribution of the $Z$-penguins shown in Fig.~\ref{fig:penguinlq} 
(the $V 2l2f$ oval of Fig.~\ref{fig:RGEs}),
not including  the  negligible effect of  the RGEs.
}

The scalar 4-lepton coefficient  $ C_{SXX}^{e\mu ee}$  can be generated via
a box diagram, with Higgs insertions on the internal quark  lines (so the coefficient can be significant for internal top quarks); however,  the coupling constant combination
that appears on the flavour-changing  line is already strictly constrained by $\meg$.
So this coefficient has a very small contribution to $\mu \to e$ processes, and we neglect it.

A classic signature of leptoquarks is $\muc$,  which can be mediated  at tree level via scalar or vector operators involving first generation quarks. The constraint from light targets like Titanium or Aluminium can be written (for outgoing $e_L$)
\bea
\sqrt{ \frac{BR^{exp}_{Ti}}{250 }}
&\gsim&  {\Big|} 0.250 C_{D,R}  (m_\mu)
+0.37  \lambda^{e u}_L \lambda^{\mu u *}_L .
\left(1{ +} \frac{{ 2}\alpha}{\pi} \ln \right)
+0.39\left(
\frac{g^2}{64\pi^2} \lambda^{e u}_L \lambda^{\mu u *}_L  \ln\frac{m_{LQ}}{m_W}\right)   \nonumber\\&&
- \frac{\alpha}{6 \pi}
  \left[\lambda_L\ln \frac{m_{LQ}}{m_Q} \lambda^\dagger_L \right]^{e\mu}  
{   - \frac{3}{{ 64}\pi^2}
\left[ \lambda_LY_u \ln \frac{m_{LQ}}{m_Q} Y_u^\dagger  \lambda^\dagger_L \right]^{e\mu} }
\label{mecTi1} \\
    && -\eta
 \left( 1.95  \lambda^{e u}_L \lambda^{\mu u *}_R 
+ \frac{0.41m_N}{27m_c}  \lambda^{e c}_L \lambda^{\mu c *}_R 
\right)
+ \eta(m_t) \frac{1.07 m_N}{27m_t}   \lambda^{e t}_L \lambda^{\mu t *}_R {\Big|}\times \frac{v^2}{m_{LQ}^2}\ , \nonumber
\eea
where  in order, the terms are: the  dipole coefficient  given in Eq.~(\ref{megLQ}),
the tree vector coefficient on $u$ quarks with its QED running
(represented as the  upper left box of Fig.~\ref{fig:RGEs}),
the electroweak box  contribution to the $d$ vector,
 the QED then  $Z$ penguin (see Fig.~\ref{fig:penguinlq})  contributions to the $u$ and $d$ vectors (where we  took  $V_{ud} = 1, \sin^2 \theta_W = 1/4$),  and the scalar $u$, $c$ and $t$ contributions. Most of the scalar top contribution comes from  a loop-induced flavour changing Higgs coupling (in SMEFT,  ${\cal O}^{e\mu tt}_{LEQU}$ mixing into  ${\cal O}^{e\mu}_{EH}$), which  generates scalar quark operators of all  quark flavours. The tensor to scalar mixing is neglected, because the model generates  tensor coefficients that  are proportional to the scalars. The  QCD and QED running of the scalars is contained in $\eta$:  
\begin{align}
\eta &= \left[\frac{\alpha_s(m_{LQ})}{\alpha_s(2{\rm GeV})}\right]^{-12/23}\!\!\!
\!\!\!\times
( 1+  \frac{13 \alpha}{ 6\pi}  \ln \frac{m_{LQ}}{2{\rm GeV}} ) \approx 1.79
,\nonumber\\
\eta(m_t)& =  \left[\frac{\alpha_s(m_{LQ})}{\alpha_s(m_t)}\right]^{-12/23} \!\!\! \!\!\!\times
( 1+  \frac{13 \alpha}{ 6\pi}  \ln \frac{m_{LQ}}{m_t} )\approx  1.11.
\end{align}

It is easy to see that the operator coefficients of Eq.~(\ref{L1}) depend on more than 12 different combinations of the leptoquark couplings,  so the only prediction of the leptoquark model is that the four-lepton scalar coefficients $C_{S,XX}^{e\mu ee}$ are negligible, as discussed after Eq.~(\ref{CVLLeeLQ});   the  model  could  fit any observed values  for  the remaining 10  coefficients.

\begin{figure}[htb]
  \begin{center}
    \includegraphics[height=8cm,width=12cm]{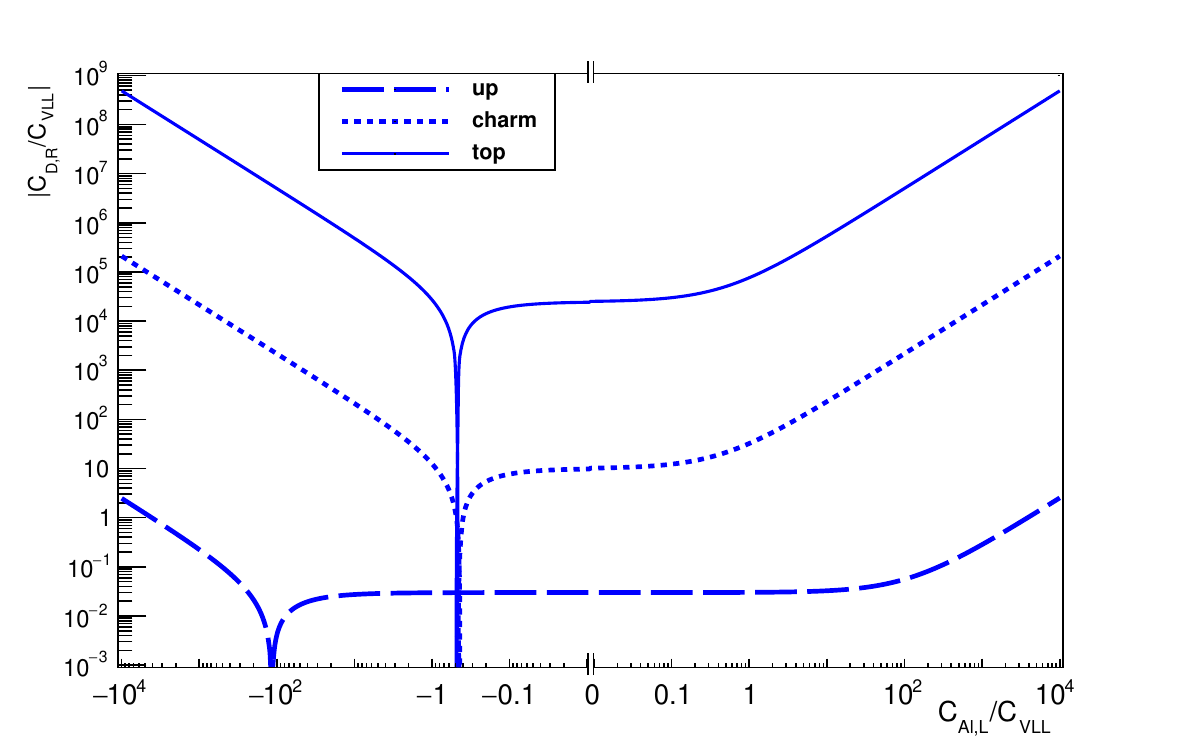}
  \end{center}
  \caption{The $S_1$  leptoquark interacting only with one quark generation   predicts relations among coefficients  given in Appendix \ref{app:LQ}:
     for large positive values of   the four-fermion coefficient on light targets  $C_{Alight,L}^{e\mu}/C_{V,LL}^{e\mu ee}$ (horizontal,{\it in log}), the ratio  $C_{DR}^{e\mu}/C_{V,LL}^{e\mu ee}$ (magnitude on the vertical)  
     is   large and positive.
   The four-lepton coefficients are  comparable  for all  quark generations.
\label{fig:LQratios} }
\end{figure}

However, the leptoquark  has the interesting feature of predicting  specific patterns of $\me$ LFV when this occurs  with only one quark generation: in this case,  all the four-lepton coefficients are of comparable magnitude,   and knowing in addition  the dipole  allows to predict $C_{AL,X}$ (or vice versa). 
To illustrate this, we neglect the box  contributions to 4-lepton operators (discussed in more detail in~\cite{pl2}), which are subdominant  for internal $t$ quarks,  but can give   an  $ {\cal O}(1)$, same-sign contribution    for internal $u$ and $c$  quarks when 
$\lambda_X^{eQ} \sim 1$.  Without the boxes, all the coefficients 
are determined by four combinations of coupling constants:    $\lambda^{eQ}_X\lambda^{\mu Q *}_X$  and  $\lambda^{eQ}_X\lambda^{\mu Q *}_Y$ for $X,Y\in\{L,R\}$ and $ Y\neq X$ (the expressions  for the coefficients are given in Appendix \ref{app:LQ}).   
This   leads to correlations among operator coefficients, as occurred in the inverse seesaw. In  Fig.~\ref{fig:LQratios}, we illustrate this correlation by plotting  ratios of coefficients, rather than the 3-$d$ plots of  Section~\ref{sec:t1} (notice that the horizontal axis  is in log$_{10}$ scale, but runs from negative to positive values, so  small values of $C_{Al,X}$ have been deleted at the origin).

One sees that ``generically'', a leptoquark coupled to the $t$ gives a large dipole, whereas a large $\muc$ rate is  expected for leptoquarks interacting with the up quark.   However,  neither of these expectations  is  an unambiguous footprint of the quark flavour dominantly coupled to the leptoquark, because $C_{Al,X}$  (resp. $C_{D,X}$) can vanish for  leptoquarks interacting with $u$ (resp. $t$) quarks.
Therefore, the observation of $\mu \to 3 e$ without $\mu \to e \gamma$ would not exclude an $S_1$ leptoquark coupling mostly to top quarks; it would just exclude generic values of the parameters of that model (i.e., values of the parameters that do not lead to cancellation in some Wilson coefficients). Similarly, the observation of $\mu \to e \gamma$ but not $\mu \to e$ conversion on light nuclei would not exclude an $S_1$ leptoquark coupling only to up quarks.

\section{Discussion and Summary}
\label{sec:summ}

In this paper, we explored whether a bottom-up EFT analysis (outlined in  Section~\ref{intro}) can give a   perspective on LFV  models  that is complementary  to  top-down studies.  We emphasize   that in EFT,   the data   for $\meg$, $\meee$ and $\muc$  consists of 
twelve  Wilson coefficients, given in Eq.~(\ref{L1}), and not just three branching  ratios.
The current experimental null-results confine the coefficients 
to the interior of a 12-D ellipse centered at the origin, and the aim of this paper was to determine whether a future observation could exclude models. To address this question, we  searched for  the regions of  coefficient space  accessible-to-future-experiments  that each model cannot reach, as an observation in that part of the  ellipse would rule the model  out.

We studied three TeV-scale\footnote{Although the EFT calculations are only logarithmically sensitive to the choice $\Lambda_{\rm NP}\sim {\rm TeV}$, our results may depend on this assumption, especially in the cases where cancellations between different contributions are envisaged.}  models: the type II seesaw, the inverse type I seesaw and a singlet scalar leptoquark added to the SM. We chose the first two because they can explain neutrino masses  (which are the best motivation for LFV), while we considered the scalar leptoquark in light of the charged current anomaly observed in  $b\to c l \nu$ transitions.
 The model predictions depend on  combinations of NP and SM parameters  which we refer to as ``invariants'', see {\it e.g.}  Eqs.~(\ref{eq:invseesawnondeg}) and~(\ref{eq:invseesawdeg}).

The type II and inverse seesaw models generate Majorana neutrino masses via the tree-level exchange of heavy new particles, respectively a scalar triplet and fermion singlets. Large lepton-flavour-changing rates are possible because the models can contain LFV
without lepton number violation, avoiding any  suppression by  small neutrino masses. In both models the new particles interact with lepton doublets, so   the coefficients of operators  with flavour-changing currents  involving singlet charged  leptons are  suppressed by the lepton Yukawa couplings\footnote{A muon Yukawa coupling also appears in $C_{DR}^{e\mu}$, but since the operator is defined with the muon mass -- see Eq.~(\ref{L1}) -- this does not count as a suppression in this case.}   and  neglected here:
  \bea
C_{D,L}\, ,\, C^{e\mu ee}_{V,RR}\, ,\, C^{e\mu ee}_{V,RL}\, ,\,
C^{e\mu ee}_{S,RR}\, ,\, C^{e\mu }_{Alight,R}\, ,\,
C^{e\mu }_{Aheavy,R}\, ,\,
C^{e\mu ee}_{S,LL}\, \simeq\, 0\nonumber
~~~~  {\rm  (type~II,~inverse~seesaw)}
\eea
 { So  in the twelve-dimensional space that can be probed by experiment, these models can only occupy 5 dimensions: should one of the above  coefficients be observed  (in the absence of the unsuppressed ones), then these  models would be excluded. In addition, these vanishing coefficients  }
imply that $\muc$ only occurs via the dipole and vector interactions.

 Section~\ref{sec:t2}  showed that
in the type II model,  three of the  remaining coefficients,
$ C^{e\mu}_{Alight, L}$, $C^{e\mu}_{Aheavy, L}$ and $ C^{e\mu ee}_{V,LR}$
(given in Eqs.~\ref{ping}, \ref{ping2})
 arise from the same loop diagrams and are  all proportional to the same combination of invariants. This implies   that the model occupies a line in the three-dimensional space of these three coefficients,  and that one of the three rates  for $\mu\to e_L  \gamma$,   $\mu\to e_L \bar e_R e_R$ and $\mu A\to e_L A$ being predicted  by the other two. 
 The  coefficients  $C^{e\mu}_{D,R}$ and $ C^{e\mu ee}_{V,LL}$ can be expressed in terms of two other   invariants,  all of which are constructed with the  Yukawa matrix of the triplet scalar.
 { This  is proportional to the  neutrino mass matrix, so known, up to  the overall magnitude,  the neutrino mass  hierarchy, the lightest  mass $m_{\rm min}$ and the Majorana Phases $\alpha_{1,2}$.}
So although the model generates  $ C^{e\mu ee}_{V,LL}$ at tree level, suggesting  $\meee$ to discover   the type II seesaw,  this coefficient can vanish
(for specific values of the Majorana phases and  a range of $m_{min}$), as could
$C^{e\mu}_{D,R}$ or  $C^{e\mu ee}_{V,LR}$.
When this occurs, the  ratio  of  the remaining two coefficients is restricted, so there are combinations of observations that the type II seesaw cannot predict.
This is illustrated in  Fig.~\ref{fig:angulart2}, where  coefficient ratios
 {(that correspond to angular coordinates in the  three remaining dimensions of the original ellipse)} are varied over the ranges  accessible to upcoming  experiments. We find  that at least one of the four-fermion coefficients is always larger than the dipole, so  that observing $\mu\to e \gamma$ with a branching ratio $Br(\mu\to e \gamma)\gtrsim 10^{-14}$ without detecting $\mu\to 3e$ in upcoming searches with $Br(\mu\to 3e)\gtrsim 10^{-16}$ can rule out the type II seesaw  { model studied here}.

Section~\ref{sec:t1} studied  the inverse seesaw model
and showed that $C^{e\mu}_{D,R}$, $C^{e\mu ee}_{V,LL}$, $C^{e\mu ee}_{V,LR}$, $C^{e\mu}_{Alight, L}$ and $C^{e\mu}_{Aheavy, L}$   are  functions of four 
invariants constructed from 
 the neutrino Yukawa and sterile neutrino mass matrices,  as given in Eq.~(\ref{eq:invseesawnondeg}).
 This  
 implies that  
 $Br(\mu Au \to e_L Au)$ could be predicted, given the rates for  $\mu Al \to e_L Al$, $\mu\to e_L \bar{e}_L e_L$, $\mu\to e_L \bar{e}_R e_R$ and $\mu \to  e_L \g$.
  The relevant contributions to these  $\mu \to e$  coefficients arise
  via loop diagrams in matching (no four-SM-fermion operators are generated at  tree-level), and are non-linear functions of the nondegenerate singlet masses. The RGEs of QED just renormalise these coefficients by a few percent, but do not generate additional invariants. 
So we observe that the number of invariants is  reduced,  if the sterile neutrino masses can be  approximated as degenerate -- as occurs for   mass differences of $\mathcal{O}(v)$, see Eq.~(\ref{degen1}).  In this limit, the 
 {five non-negligeable coefficients } are controlled by two invariants  constructed from the neutrino Yukawa matrices : $\mathcal{O}(Y_\nu Y_\nu^\dagger)$ and $\mathcal{O}(Y_\nu Y_\nu^\dagger Y_\nu Y_\nu^\dagger)$.
This implies that when two coefficients are known, the
 {remaining three are} 
predicted, implying, {\it eg}, that the model predicts $Br(\muc)$, from the rates for $\meg$ and $\meee$.  {In the twelve-dimensional ellipse, our
 inverse seesaw model  with degenerate sterile neutrinos therefore occupies a two-dimensional subspace (see Eqs.~\ref{eq:typeIcvlr} and~\ref{eq:typeIcvll}), which we } 
illustrate
in Fig.~\ref{fig:typeIcv} by plotting the model prediction for the real part of three coefficients.

Finally, in Section~\ref{sec:lq}, we investigated the $\mu\to e$ predictions of a singlet scalar leptoquark, selected to fit the   excess of $b\to c \bar{\tau} \nu$ events observed  in the $R_{D}$ ratio.
The model contributes to all but two of the $\mu\to e$ observable coefficients with different coupling combinations, implying that it could entirely fill 10 dimensions of the 12-D ellipse. Only the observation of a non-zero $\mu\to 3e$ scalar coefficient $C^{e\mu ee}_{S,XX}$ could not be explained by the leptoquark. On the other hand, the model is more predictive
when the leptoquark only interacts with one quark generation.
In this case,  all the invariants become $\propto \lambda_X^{eQ} \lambda_L^{ \mu Q *}$ or $\lambda_X^{eQ} \lambda_R^{ \mu Q *}$, so once  four  coefficients are measured,  the  remaining eight can be predicted.   For a specific chirality of the  outgoing electron  in the LFV current, this resembles  the degenerate inverse seesaw case, and the equations relating the coefficients are given in Appendix \ref{app:LQ}.   
The relations between coefficient ratios  that are expected when the leptoquark only interacts with  one quark generation  are illustrated  in Fig. \ref{fig:LQratios}.

The results of this paper will be extended in a subsequent publication, where also some technical details of our EFT calculations will be discussed. We will explore the impact of complementary observables {and the uses of invariants},  and we will discuss the consequences of relative complex phases for the operator coefficients.

In summary, we find that there are  observations of $\mu \to e$ processes that could rule out the three models we considered.
The type II seesaw model predicts coefficients in  part of a 3-dimensional subspace of the 12-d coefficient space accessible to experiments. The  inverse seesaw   maps  onto a 4-d subspace of the 12-d  space, in the case of  non-dengenerate sterile neutrinos, but  is more predictive for  (nearly) degenerate steriles, where it  is restricted to a 2-dimensional subspace. The singlet scalar leptoquark model does not generate sizable scalar four-lepton operators but  can give arbitrary contributions to all other Wilson coefficients, thus completely filling 10 dimensions of the 12-d ellipse. However, if the leptoquark couplings to the electron and muon involve a single quark generation, the model predictions are restricted to a 4-dimensional subspace.

\subsubsection*{Acknowledgements}
We thank Luca Silvestrini for a helpful suggestion, and Ann-Kathrin Perrevoort for helpful comments about observables in $\meee$.  MA was supported by a doctoral fellowship from the IN2P3.
The work of SL is supported in part by the European Union's Horizon 2020 research and innovation programme under the Marie Sklodowska-Curie grant agreement No. 860881-HIDDeN.

\appendix


\section{Appendix: Branching Ratios}
\label{app:BRs}
For completeness, we list here the branching ratios for $\meg$, and  $\meee$:
\bea
BR(\meg)&=& 384 \pi^2 (|C^{e\mu}_{D L}|^2 + |C^{e\mu}_{D R}|^2) ~~~, 
\label{BRmeg}\\
BR(\meee) 
& =&  \frac{|C^{e\mu ee}_{S,LL}|^2+ |C^{e\mu ee}_{S,RR}|^2}{8}
 + (64 \ln\frac{m_\mu}{m_e} -136) 
 (|eC^{e\mu}_{D ,R}|^2 +|eC^{e\mu}_{D ,L}|^2)
\nonumber\\
&&
+2 |C^{e\mu ee}_{V,RR}  + 4eC^{e\mu}_{D, L}|^2
 +2| C^{e\mu ee}_{V,LL}  + 4eC^{e\mu}_{D ,R}|^2 
\label{BRmeee}\\
&&
 + |C^{e\mu ee}_{V,RL}  + 4eC^{e\mu}_{D, L}|^2
 + |C^{e\mu ee}_{V,LR}  + 4eC^{e\mu}_{D, R} |^2 ~~~.
 \nonumber
\eea

\section{Appendix: the $\muc$ operators}
\label{app:mucops}

 The Spin-Independent $\mu\to e$  conversion rate, normalised to the $\mu$ capture rate~\cite{KKO,Suzuki:1987jf}  can be written~\cite{KKO}
    \bea
{\rm BR}_{SI}(\mu A \to eA) &=&   \frac{32G_F^2 m_\m^5 }{ \Gamma_{cap}}   
 {\Big [ } \big|     
   \widetilde{C}^{pp}_{V,R} I_{A,V}^{(p)} + \widetilde{C}^{pp}_{S,L}  I_{A, S}^{(p)}
+ \widetilde{C}^{nn}_{V,R} I_{A, V}^{(n)} + \widetilde{C}^{'nn}_{S,L}  I_{A,S}^{(n)} 
+  C_{D,L} {\frac{I_{A,D}}{4}}  
 \big|^2  \nonumber \\ &&+ \{ L \leftrightarrow R \}~ {\Big ]} ~~~,
 \label{BRmecKKO}
 \eea
 where   $ I_{A,V}^{(N)}$, $I_{A,S}^{(N)} $ and $I_{A,D}$
are  target($A$)-dependent 
``overlap  integrals''  inside the nucleus of the lepton wavefunctions
and the appropriate  nucleon  density.
This shows that    a target probes a linear combination of coefficients  
identified by the overlap integrals.  With  current theoretical uncertainties  on the overlap integrals,  at least two  independent combinations of  the coefficients-on-nucleons $\{\tilde{C}\}$  could be constrained~\cite{DKY}.  We will  take these two combinations to correspond to light and heavy nuclei.

For light targets like Aluminium or Titanium, all the four-fermion overlap integrals are comparable, so the  four-fermion operator that is probed is approximately 
\bea
\mathcal{O}_{Alight,X} \approx \frac{1}{2}{\Big (}  (\overline{e} P_X \mu)  (\overline{p} p)
+(\overline{e}\g^\a P_X \mu)  (\overline{p}\g_\a p)
+(\overline{e} P_X \mu)  (\overline{n}  n)
+  (\overline{e}\g^\a P_X \mu)  (\overline{n}\g_\a n)
  {\Big) } \hskip .5cm
  \eea
  or more precisely, the  KKO calculation says that the combination of  coefficients probed by Aluminium 
  is~\cite{DKY}
\bea
\tilde{C}_{Alight,X} &=& 0.455 \widetilde{C}^{pp}_{S,X}
 +0.473 \widetilde{C}^{pp}_{V,Y} + 0.490\widetilde{C}^{nn}_{S,X}
 +0.508 \widetilde{C}^{nn}_{V,Y}) ~~~.
 \nonumber
 \eea
 For Gold,  the coefficient combination is slightly misaligned:
 \bea
 \tilde{C}_{Aheavy,X}& =& 0.289 \widetilde{C}^{pp}_{S,X}+ 0.458\widetilde{C}^{pp}_{V,Y} + 0.432\widetilde{C}^{nn}_{S,X} +0.686 \widetilde{C}^{nn}_{V,Y}~~~, \nonumber
 \eea
 indeed, the operator probed by heavy targets can be written as
$\mathcal{O}_{Aheavy,X}  = \cos\phi  \mathcal{O}_{Alight,X} + \sin \phi  \mathcal{O}_{Aheavy\perp,X}  $.
 Measuring the coefficient  of $ \mathcal{O}_{Aheavy\perp,X} $ is the new information that can be obtained from heavy targets, but not light ones.

 The definition of $ \mathcal{O}_{Aheavy\perp,X}  $  depends on whether it is constructed in the nucleon EFT, or the  quark EFT relevant above a scale of 2 GeV. 
This is because    there is information loss in  matching nucleons to quarks, because  the scalar densities of both  $u$ and $d$ quarks in the  neutron and  proton are all comparable, so   the scalar $u$ and $d$ coefficients $C_{SX}^{qq}$ are indistinguishable 
unless the scalar nucleon coefficients $ \widetilde{C}^{NN}_{S,X}$ are accurately measured. In addition,  there is a several-$\sigma$ discrepancy between the
determinations of the scalar quark densities in the nucleon from the lattice and  pion data.

So in this paper we focus on  $\muc$ on light targets, because only 
the leptoquark induces scalar quark  coefficients,   and we prefer to avoid the   quark-scalar uncertainties associated with defining 
 $ \mathcal{O}_{Aheavy\perp,X}  $ .  We will consider the complementary information from heavy targets in~\cite{pl2}.

\section{Appendix: If the leptoquark interacts only with one generation of quarks}
\label{app:LQ}

In this appendix, we give formulae for the operator coefficients in the leptoquark model, for the case where the leptoquak interacts only with one generation of quarks. 

If the leptoquark only interacts with top quarks,
    one obtains:
\bea
 \frac{ m_{LQ}^2}{ v^2}C_{DR} (m_\mu) &\simeq& 
2.3\times 10^{-4} [\lambda_L\lambda^\dagger_L]_{e\mu} 
-12 [\lambda_L  \lambda^\dagger_R]_{e\mu}  \label{megLQt}
\\
 \frac{ m_{LQ}^2}{ v^2} C_{V,XY}^{e\mu ee} (m_\mu) & \simeq &
-4.45\times 10^{-3}  [\lambda_X \lambda_X^\dagger]_{e \mu} [\lambda_Y \lambda_Y^\dagger]_{ee}\nonumber\\
&{ +}& (1.36\times 10^{-3}  - g^e_Y 0.033) [\lambda_X  \lambda^\dagger_X]_{e\mu}
\\
\frac{m_{LQ}^2}{v^2} C_{Alight,L} &\simeq& -  
   6.8\times 10^{-4}
  \left[\lambda_L \lambda^\dagger_L \right]^{e\mu}  
-0.0084
[\lambda_L  \lambda^\dagger_L ]^{e\mu}  
{+} 2.4 \times 10^{-4}  \lambda^{e t}_L \lambda^{\mu t *}_R  \nonumber
\eea
where  $X,Y \in \{L,R\}$ for the vector four-lepton  coefficients, for which  the contributions   due to   boxes are included.
The three terms of the  four-fermion contribution  to $\muc$  are induced by the photon and $Z$ penguins, and the top-loop contribution to  the  $\me$  Yukawa coupling.

   If the leptoquark only interacts with charm quarks, then:
\bea
 \frac{ m_{LQ}^2}{ v^2}C_{DR} (m_\mu) &\simeq& 
2.3\times 10^{-4} [\lambda_L\lambda^\dagger_L]_{e\mu} 
-0.42
 [\lambda_L  \lambda^\dagger_R]_{e\mu} 
 \label{megLQc}
 \\
 \frac{ m_{LQ}^2}{ v^2} C_{V,XY}^{e\mu ee} (m_\mu) & \simeq &
 -4.5\times 10^{-3}  [\lambda_X  \lambda^\dagger_X]_{e\mu} [\lambda_Y \lambda_Y^\dagger]_{ee}
{+} (4.8\times 10^{-3} -g^e_Y 1.9\times 10^{-6})  [\lambda_X  \lambda^\dagger_X]_{e\mu}
 \nonumber\\
 \frac{m_{LQ}^2}{v^2} C_{Alight,L} &\simeq& -
   2.4\times 10^{-3}   \left[\lambda_L \lambda^\dagger_L \right]^{e\mu}  
{-} 0.02  \lambda^{e c}_L \lambda^{\mu c *}_R  \nonumber
  \eea
where $X,Y \in \{L,R\}$ in the vector four-lepton coefficients,
for which   the $Z$-penguin ( the last term) is suppressed $\propto m_c^2/v^2$, 
and  any box can contribute   because    $(g-2)_e$ only  constrains 
 $  \lambda^{ec}_L  \lambda^{ec *}_R < 0.7$.

And finally for  a leptoquark that only has  $\me$ interactions on $u$ quarks, one obtains:
\bea
 \frac{ m_{LQ}^2}{ v^2}C_{DR} (m_\mu) &\simeq& 
 2.3\times 10^{-4} [\lambda_L\lambda^\dagger_L]_{e\mu}
 - 7.3\times 10^{-4}
 [\lambda_L  \lambda^\dagger_R]_{e\mu} 
 \label{megLQu}
 \\
 \frac{ m_{LQ}^2}{ v^2} C_{V,XY}^{e\mu ee} (m_\mu) & \simeq &
- 4.5\times 10^{-3}  [\lambda_X  \lambda^\dagger_X]_{e\mu} [\lambda_Y \lambda_Y^\dagger]_{ee}
{+} 4.8\times 10^{-3}   [\lambda_X  \lambda^\dagger_X]_{e\mu}
 \nonumber\\
 \frac{m_{LQ}^2}{v^2} C_{Alight,L} &\simeq& 
0.38 \lambda^{e u}_L \lambda^{\mu u *}_L 
\label{mecTiu} 
- 2.0 \eta    \lambda^{e u}_L \lambda^{\mu u *}_R 
\eea

\end{document}